\documentclass[aps,prd,longbibliography,twocolumn,preprintnumbers,nofootinbib,
superscriptaddress,amsmath,floatfix]{revtex4}

\usepackage{amssymb}  
\usepackage{graphicx,subfigure}
\usepackage{exscale}
\usepackage{textcomp}
\usepackage{enumerate}
\usepackage [english]{babel}
\usepackage[utf8]{inputenc}
\usepackage [autostyle, english = american]{csquotes}
\MakeOuterQuote{"}

\usepackage{color}

\usepackage[normalem]{ulem}  

\newcommand{\non}{\nonumber\\}

\newcommand{\be}{\begin{equation}}
\newcommand{\ee}{\end{equation}}
\newcommand{\bea}{\begin{eqnarray}}
\newcommand{\eea}{\end{eqnarray}}
\newcommand{\ba}[1]{\begin{array}{#1}}
\newcommand{\ea}{\end{array}}

\newcommand{\bm}[1]{\mbox{\boldmath${#1}$}}

\begin{document}

\title{Surface tension of dense matter at the chiral phase transition}

\author{Eduardo S.\ Fraga}
\email{fraga@if.ufrj.br}
\affiliation{Instituto de F\'{\i}sica, Universidade Federal do Rio de Janeiro, 
Caixa Postal 68528, 21941-972, Rio de Janeiro, RJ, Brazil}

\author{Maur\'{\i}cio Hippert}
\email{hippert@if.usp.br}
\affiliation{Instituto de F\'{\i}sica, Universidade Federal do Rio de Janeiro, 
Caixa Postal 68528, 21941-972, Rio de Janeiro, RJ, Brazil}
\affiliation{Instituto de F\'{\i}sica, Universidade de  S\~{a}o Paulo,  Rua  do  Mat\~{a}o, 1371,  Butant\~{a},  05508-090,  S\~{a}o  Paulo,  SP,  Brazil}

\author{Andreas Schmitt}
\email{a.schmitt@soton.ac.uk}
\affiliation{Mathematical Sciences and STAG Research Centre, University of Southampton, 
Southampton SO17 1BJ, United Kingdom}

\date{23 January 2019}

\begin{abstract} 
If a first-order phase transition separates nuclear and quark matter at large baryon density, 
an interface between these two phases has a nonzero surface tension. We calculate this surface tension within a nucleon-meson model for domain walls and bubbles. Various methods and approximations are discussed and compared, including a numerical evaluation of the spatial profile of the interface. We also compute the surface tension at the other first-order phase transitions of the model: the nuclear liquid-gas transition and, in the parameter regime where it exists, the direct transition from the vacuum to the (approximately) chirally symmetric phase. Identifying the chirally symmetric phase with quark matter -- our model does not contain explicit quark degrees of freedom -- we find maximal surface tensions of the vacuum-quark transition $\Sigma_{\rm VQ}\sim 15 \, {\rm MeV}/{\rm fm}^2$, relevant for the surface of quark stars, and of the nuclear-quark transition $\Sigma_{\rm NQ}\sim 10 \, {\rm MeV}/{\rm fm}^2$, relevant for hybrid stars and for quark matter nucleation in supernovae and neutron star mergers. 

\vspace{1.5cm}
\end{abstract}

\maketitle

\section{Introduction}

Strong interactions, and possibly quark matter, play a crucial role in the astrophysics of neutron stars \cite{glendenningbook,Schmitt:2010pn}, from the structure of their cores to cataclysmic events associated with their formation in supernovae explosions and collision in neutron star mergers \cite{Pons:2001ar,Oechslin:2004yj,Sagert:2008ka,Fischer:2010wp,Kurkela:2010yk,Paschalidis:2017qmb,Fischer:2017lag,Most:2018eaw,Christian:2018jyd,Bauswein:2018bma,Cao:2018tzm,Han:2018mtj}. 

To assess the possibility of the formation and evolution of quark matter in such systems one needs, besides the knowledge of the equation of state for nuclear matter under such extreme densities, information on the relevant time scales. At sufficiently high energy densities, one expects strongly interacting matter to become deconfined and essentially chiral \cite{Rischke:2003mt,Stephanov:2007fk,Fukushima:2010bq}, and thus chiral quark matter could provide the relevant degrees of freedom in the core of compact stars \cite{Weber:2004kj,Alford:2006vz}. In fact, it was shown that deconfinement can happen at an early stage of a core-collapse supernova process, which could result not only in a delayed explosion but also in a neutrino signal of the presence of quark matter in compact stars \cite{Sagert:2008ka}. However, the issue depends dramatically on the time scales of phase conversion as compared to the time during which the superdense region is probed \cite{Mintz:2009ay,Bombaci:2009jt}. 

Most model descriptions of strong interactions at high density and low temperature suggest a first-order phase transition for the chiral and the deconfinement transitions. If that is the case, a key ingredient is the surface tension, which sets the time scale -- together with the pressure difference at the phase transition and the temperature -- for the phase conversion process \cite{Mintz:2009ay}. Indeed, the surface tension is relevant for bubble nucleation of quark matter in supernovae \cite{Sagert:2008ka,Bombaci:2009jt,Mintz:2009ay,Mintz:2010mh,Logoteta:2012ms,Jimenez:2017fax} and neutron star mergers
\cite{Chen:2013zya,Most:2018eaw}. It is also relevant for a possible quark-hadron mixed phase in the interior of neutron stars \cite{Glendenning:1992vb,Heiselberg:1992dx,Voskresensky:2002hu}. In this case, Coulomb effects together with the surface tension determine whether a mixed phase -- being electrically neutral globally, but not locally -- is preferred in a certain parameter regime over a globally {\it and} locally neutral homogeneous phase. 

Ideally, the surface tension should be calculated from the underlying fundamental theory, Quantum Chromodynamics (QCD). However, current first-principle methods are not sufficient to determine the phase structure at large baryon chemical potentials, let alone to compute the surface tension at a possible first-order chiral or deconfinement phase transition. The only available rigorous methods for dense QCD are essentially effective theories for nuclear matter at relatively low baryon density and perturbative calculations for quark matter at ultra-high baryon density.  The density regimes where the two approaches are valid are far apart, such that at a possible first-order transition at least one of them, very likely both, cannot be trusted \cite{Kurkela:2014vha}. 
Even if they happened to separately describe the low- and high-density phases reasonably well near the transition, this would be of little use for a rigorous calculation of the surface tension.  
The reason is that calculating the surface tension requires the knowledge of the entire potential, i.e., a single approach that contains nuclear and quark matter is required. At large baryon density, no such approach within QCD is currently available. Therefore, we currently rely on simple estimates or model calculations that contain both phases in a more or less realistic way \cite{PhysRevD.30.2379,PhysRevC.35.213,PhysRevLett.70.1355,Alford:2001zr,Palhares:2010be,Lugones:2011xv,Palhares:2011jd,PhysRevD.84.036011,Pinto:2012aq,Mintz:2012mz,Lugones:2013ema,Ke:2013wga,Gao:2016hks}\footnote{The surface tension {\it can} be calculated from first principles within lattice QCD
if quarks are assumed to be unphysically heavy such that there is a first-order phase transition at zero chemical potential \cite{KAJANTIE1991693,PhysRevD.42.2864,BHATTACHARYA1992497,Beinlich:1996xg,Lucini:2005vg}. }.   

Previous estimates of the surface tension were either performed in the framework of chiral models that lack the nuclear matter ingredient or employed two different models for nuclear and quark matter, which are glued together at the phase transition. To fill this gap, we employ a nucleon-meson model \cite{Boguta:1982wr,Boguta:1986ha,Floerchinger:2012xd,Drews:2013hha} that contains realistic nuclear matter, in the sense that its parameters are matched to the known properties of nuclear matter at saturation density. 
It does not contain quark degrees of freedom, thus the chirally restored phase can only be considered as a very rough approximation of dense quark matter. 

Our model exhibits three, not just two, potentially stable phases: vacuum, nuclear matter, and the phase where the approximate chiral symmetry is restored ("quark matter"). As a consequence, there are three possibilities for first-order phase transitions. 
Besides transitions from the vacuum to nuclear matter and from nuclear to quark matter, a direct transition from the vacuum to quark matter is possible, depending on the parameter set.  
We identify the parameter regimes for these possibilities and compute the surface tension for all three transitions. The vacuum-quark surface tension is of phenomenological interest for the crust of strange stars \cite{Alcock:1986hz,Alford:2008ge} or so-called "strangelet dwarfs" which may exist if the surface tension is sufficiently small \cite{Alford:2011ue}. 

Finally, we also discuss various methods of calculating the surface tension. This is of particular interest because our model contains two mesonic condensates which have to be determined dynamically. This renders the calculation more complicated compared to the case of a single condensate. We compare numerical calculations using the domain wall and bubble profiles with approximations that merely require a numerical integration rather than solving a system of differential equations. 

Our paper is organized as follows. In Sec.\ \ref{sec:model} we introduce the model and the approximations we use. We discuss the homogeneous phases in Sec.\ \ref{sec:hom}, as a necessary preparation for calculating the surface tension. The computation of the surface tension is divided into two sections. In Sec.\ \ref{sec:calc} we explain the calculation for planar and spherical geometries, present several ways to approximate the full numerical result, and show some selected results, including temperature effects. 
In Sec.\ \ref{sec:results} we give a complete survey of the zero-temperature results for the various surface tensions in the parameter space of the model. Sec.\ \ref{sec:summary} presents our summary and outlook.

\section{Setup}
\label{sec:model}

\subsection{Model Lagrangian and approximations}

To model nuclear matter and chiral symmetry at low temperatures and high densities, we adopt the following Lagrangian \cite{Boguta:1982wr,Boguta:1986ha,Floerchinger:2012xd,Drews:2013hha},
\bea \label{Lmod}
{\cal L} &=& 
\bar{\psi}(i\gamma_\mu\partial^\mu + \gamma^0\mu)\psi + \frac{1}{2}\left(\partial_\mu\sigma\partial^\mu\sigma + \partial_\mu\bm{\pi}\cdot\partial^\mu\bm{\pi}\right) \non[2ex]
&&- \sum_{n=1}^4 \frac{a_n}{n!} \frac{(\sigma^2+\bm{\pi}^2-f_\pi^2)^n}{2^n}+\epsilon(\sigma-f_\pi) \non[2ex]
&&-\frac{1}{4}\omega_{\mu\nu}\omega^{\mu\nu}+\frac{1}{2}m_\omega^2\omega_\mu\omega^\mu \non[2ex]
&&- g_\sigma\bar{\psi}(\sigma+ i\gamma^5\bm{\tau}\cdot\bm{\pi})\psi - g_\omega\bar{\psi}\gamma_\mu\omega^\mu\psi \, , 
\eea
where $\psi$ is the nucleon spinor, and $\sigma$, $\bm{\pi}$, $\omega_\mu$ are the mesonic fields. We have 
denoted $\omega_{\mu\nu}\equiv \partial_\mu\omega_\nu-\partial_\nu\omega_\mu$, $\bm\tau=(\tau_1,\tau_2,\tau_3)$ are the Pauli matrices, and $\mu$ is the baryon chemical potential. The nucleon spinor has four degrees of freedom in Dirac space and two degrees of freedom in isospin space, corresponding to neutrons and protons. 
We restrict ourselves to isospin-symmetric nuclear matter, such that these two degrees of freedom are degenerate. The pion decay constant and the omega mass are given by $f_\pi\simeq 93\,{\rm MeV}$ and  $m_\omega\simeq 782\,{\rm MeV}$, respectively. The model contains a mesonic potential with 5 parameters $a_1, \ldots, a_4, \epsilon$, and Yukawa interactions between the nucleons and the mesons with the coupling constants $g_\sigma$ and $g_\omega$. The matching of these 7 parameters will be discussed below. For $\epsilon=0$, the Lagrangian is chirally symmetric. 
A (small) nonzero $\epsilon$ introduces a (small) explicit symmetry breaking, while condensation of the $\sigma$ field breaks the (approximate) chiral symmetry spontaneously. Notice that the mass of the nucleon is generated entirely by chiral symmetry breaking, since there is no nucleon mass parameter in the Lagrangian. As mentioned in the introduction, the choice of the model is mainly 
motivated by the possibility to simultaneously account for realistic nuclear matter and for a chirally symmetric phase, and by its relative simplicity, allowing for a thorough 
calculation of the surface tension. A similar choice, albeit leading to more complicated calculations, would be the extended linear sigma model from 
Refs.\ \cite{Detar:1988kn,Gallas:2011qp,Haber:2014ula}. Another option, 
of particular importance for the application to neutron stars, is the extension of the present model to isospin-asymmetric nuclear matter \cite{Drews:2014spa}. 

In what follows, we employ the mean-field approximation, i.e., we neglect mesonic fluctuations. These are expected to become particularly important at large temperatures, whereas we work only at zero or small temperatures. Moreover, we work within the "no-sea" approximation, i.e., we simply drop the vacuum term in the pressure, while a more elaborate 
evaluation would include renormalization, with the vacuum contribution depending on the renormalization scale. This would not render the calculation much more difficult and might be considered more rigorous from a theoretical point of view. However, the 
model is phenomenological to begin with, so it is unclear whether the result of a more complete evaluation is more realistic in any sense. Finally, we make use of the Thomas-Fermi approximation when computing the surface tension. This approximation has been employed frequently in the literature (e.g., Refs.\ \cite{Menezes:1999zz,PhysRevD.84.036011}) and simplifies the calculation tremendously. In this approximation, 
the spatial dependence of the mesonic condensates 
is ignored when the fermions are integrated out, as opposed to a full solution of the Dirac equation for the fermions. Strictly speaking, this approximation 
is only valid for small gradients of the condensates. We shall nevertheless explore the full parameter space, including regions with large gradients, having 
in mind that our approximation has to be taken with care in these regions. These approximations lead to a relatively simple setup as follows.   

The mesonic condensates $\bar{\sigma}$ and $\bar{\omega}_\mu$ are introduced via the usual shift $\sigma\to\bar{\sigma}+\sigma$ and 
$\omega_\mu\to \bar{\omega}_\mu +\omega_\mu$, where now $\sigma$ and $\omega_\mu$ are fluctuations. We do not include the possibility of omega condensation in the spatial components, which would break rotational symmetry, $\bar{\omega}_i=0$, and then omit the subscript $0$ from the omega condensate for simplicity, $\bar{\omega}\equiv \bar{\omega}_0$.  After separating the condensates, the Lagrangian becomes  
\bea \label{Lmf}
{\cal L} &=& \bar{\psi}(i\gamma_\mu\partial^\mu -M + \gamma^0\mu_*)\psi  +\frac{(\nabla \bar{\omega})^2}{2}+\frac{1}{2}m_\omega^2\bar{\omega}^2 \non[2ex]
&&- \frac{(\nabla \bar{\sigma})^2}{2} - U(\bar{\sigma}) + \mbox{fluctuations} \, ,
\eea
where we have assumed the condensates to be time-independent but kept their spatial dependence. Notice that, being analogous to the Coulomb field in electrodynamics, $\omega_0$ -- and thus $\bar\omega$ -- is not an independent dynamical field; hence the different sign in front of its gradient 
contribution. Its Euler-Lagrange equation, which we shall use below, has 
thus to be understood as a constraint rather than a minimization of the free energy.
The fluctuation terms, containing $\sigma$, $\omega_\mu$ and $\bm{\pi}$, will be 
neglected from now on. As customary, we have introduced the 
effective nucleon mass and the effective baryon chemical potential, 
\be 
M \equiv  g_\sigma \bar{\sigma}\, , \qquad \mu_* \equiv \mu-g_\omega\bar{\omega} \, ,
\ee
and abbreviated the potential for the $\sigma$ condensate,
\be 
U(\bar{\sigma}) = \sum_{n=1}^4 \frac{a_n}{n!} \frac{(\bar{\sigma}^2-f_\pi^2)^n}{2^n} - \epsilon(\bar{\sigma}-f_\pi) \, .
\ee
The sigma and pion masses are read off from the quadratic terms in $\sigma$ and $\bm{\pi}$, not shown explicitly in Eq.\ (\ref{Lmf}).  In the vacuum, requiring $\bar{\sigma}=f_\pi$ yields $m_\sigma^2 = a_1 + f_\pi^2 a_2$ and $m^2_\pi = a_1$. With the pion mass $m_\pi=139\, {\rm MeV}$, this fixes the parameter $a_1$. The constraint $\bar{\sigma}=f_\pi$ can also be used to fix $\epsilon$: $f_\pi$ is a minimum of $U(\bar{\sigma})$ if $\epsilon=m_\pi^2 f_\pi$. We fix the scalar coupling through $g_\sigma = m_N/f_\pi \simeq 10.097$ with the vacuum mass of the nucleon 
$m_N=939\,{\rm MeV}$. We are thus left with the 4 parameters $g_\omega, a_2, a_3, a_4$, whose matching to nuclear matter properties we discuss 
in Sec.\ \ref{sec:fix}. 

Now, integrating over the fermionic fields within the Thomas-Fermi approximation and dropping the vacuum contribution, we arrive at the free energy 
density
\bea \label{Omega}
\Omega 
&=&-\frac{(\nabla\bar{\omega})^2}{2}+\frac{(\nabla\bar{\sigma})^2}{2}+\Omega_0(\bar{\sigma},\bar{\omega}) \, , 
\eea
where we have separated the gradient terms, such that $\Omega_0$ depends only on the condensates themselves, not their derivatives, 
\bea\label{Omega0}
\Omega_0(\bar{\sigma},\bar{\omega}) 
&\equiv& -\frac{1}{2}m_\omega^2\bar{\omega}^2+U(\bar{\sigma}) \non[2ex]
&&- 4T\sum_{e=\pm} \int\frac{d^3{\bf k}}{(2\pi)^3} \ln\left(1+e^{-\frac{\varepsilon_{k}-e \mu_*}{T}}\right) \, , 
\eea
where $T$ is the temperature and $\varepsilon_k=\sqrt{k^2+M^2}$ is the single-nucleon energy. The factor $4$ accounts for the two spin and two isospin degrees of freedom, which are all degenerate. We have kept the anti-nucleon contribution, $e=-$, although it is negligibly small for the temperature regime we are interested in. The Euler-Lagrange equations for the meson condensates are then
\begin{subequations}\label{dOm}
\bea
\nabla^2 \bar{\sigma} &=& \frac{\partial \Omega_0}{\partial\bar{\sigma}}
= \frac{\partial U}{\partial\bar{\sigma}} + g_\sigma n_s \, ,  \label{eom1} \\[2ex] 
\nabla^2 \bar{\omega} &=& 
-\frac{\partial\Omega_0}{\partial\bar{\omega}} = m_\omega^2\bar{\omega} - g_\omega n_B  \, , \label{eom2}
\eea
\end{subequations} 
where $n_s$ and $n_B$ are scalar and baryon densities,
\begin{subequations}
\bea
&&n_s =  4M \sum_{e=\pm}\int\frac{d^3{\bf k}}{(2\pi)^3} \frac{f(\varepsilon_{k}-e\mu_*)}{\varepsilon_k}\non[2ex]
&&\stackrel{T=0}{\longrightarrow} \;
\Theta(\mu_*-M) \frac{M}{\pi^2}\left(k_F\mu_*-M^2\ln\frac{k_F+\mu_*}{M}\right) , \\[2ex] 
&&n_B =  4 \sum_{e=\pm} e\int\frac{d^3{\bf k}}{(2\pi)^3} f(\varepsilon_{k}-e\mu_*) \non[2ex]
&&\stackrel{T=0}{\longrightarrow} \;\Theta(\mu_*-M) \frac{2k_F^3}{3\pi^2} \, , \label{nB}
\eea
\end{subequations}
with the Fermi distribution $f(x) = (e^{x/T}-1)^{-1}$ and the nucleon Fermi momentum $k_F = \sqrt{\mu_*^2-M^2}$.

\subsection{Matching parameters to nuclear matter at saturation}
\label{sec:fix}

To fix the remaining parameters $g_\omega, a_2, a_3, a_4$, we make use of the following properties of infinite, zero-temperature, symmetric nuclear matter at saturation,
\bea \label{n0E} 
n_0 &=& 0.153 \, {\rm fm}^{-3} \, , \quad  E_{\rm bind} = -16.3 \, {\rm MeV} \, , \non[2ex]
M_0 &\simeq& (0.7-0.8) m_N \, , \quad K \simeq (200 - 300) \, {\rm MeV} \, ,
\eea
where the saturation density $n_0$ and the binding energy $E_{\rm bind}$ are known to good accuracy, while for the effective Dirac mass $M_0$ and the incompressibility $K$ at saturation only a certain range is known (see, e.g., Refs.\ \cite{1989NuPhA.493..521G,Johnson:1987zza,Li:1992zza,glendenningbook,Jaminon:1989wj,Furnstahl:1997tk} and \cite{Blaizot:1995zz,Youngblood:2004fe}, respectively). Of course, the values we use here for the bounds of these ranges 
contain some arbitrariness and should not be taken too literally. 

We first notice that the calculation of $g_\omega$ decouples:  from the  definition of $\mu_*$, we know that $g_\omega\bar{\omega}=\mu_0-\sqrt{k_F^2+M_0^2}$  at saturation, where the chemical potential is given by $\mu_0=m_N+E_{\rm bind}=922.7\, {\rm MeV}$ and the Fermi momentum $k_F$ is expressed in terms of $n_0$ with the help of Eq.\ (\ref{nB}). From Eq.\ (\ref{eom2}), evaluated 
in the homogeneous case, $\nabla^2 \bar{\omega} = 0$, we obtain $m_\omega^2\bar{\omega} = g_\omega n_0$. Putting this together yields 
\be
g_\omega^2 = \frac{m_\omega^2}{n_0}\left[\mu_0-\sqrt{\left(\frac{3\pi^2 n_0}{2}\right)^{2/3}+M_0^2}\right] \, .
\ee
Hence for a given $M_0$ (and given $n_0$ and $E_{\rm bind}$) we can directly compute $g_\omega$ without having to choose a value for $K$. Next, $a_2$, $a_3$, $a_4$ are computed from the following three coupled equations: Eq.\ (\ref{eom1}) with $\nabla^2 \bar{\sigma}=0$; the condition that at the nuclear matter onset the free energy $\Omega$ of  nuclear matter be identical to that of the vacuum, $\Omega=0$; 
and the condition that the incompressibility $K$ be given by a chosen value, where the incompressibility can be written as \cite{glendenningbook}
\bea \label{comp}
&&K = 9n_B\frac{\partial^2\varepsilon}{\partial n_B^2} = \frac{6k_{F}^3}{\pi^2} \left(\frac{g_\omega}{m_\omega}\right)^2 + \frac{3k_{F}^2}{\mu_*} \non[2ex]
&&-\frac{6k_{F}^3}{\pi^2}
\left(\frac{M}{\mu_*}\right)^2\left[\frac{\partial^2 U}{\partial M^2}+\frac{2}{\pi^2}\int_0^{k_{F}}dk\,\frac{k^4}{\varepsilon_k^3}\right]^{-1} ,
\eea
where $\varepsilon$ is the energy density. 

Additionally, one may use the surface tension of nuclear matter at saturation, 
\be \label{Sigma0}
\Sigma_0 \simeq (1.0 - 1.2) \, {\rm MeV} \, {\rm fm}^{-2} ,
\ee
as a constraint for the parameter space \cite{Hua:2000gd,Floerchinger:2012xd,Drews:2013hha}. Again, the range of this quantity is more or less known, but the numbers chosen here for the upper and lower boundaries are somewhat arbitrary. 
Even though this constraint introduces a 5th quantity for matching the 4 parameters $g_\omega, a_2, a_3, a_4$, it is conceivable that all constraints are met because $K$, $M_0$, and $\Sigma_0$ are only known within a nonzero range. 
However, in contrast to the other 4 saturation properties, the surface tension is prone to errors if computed from the given Thomas-Fermi approximation, and thus it has to be considered with some care if treated as a further constraint. In fact, we shall show later that, if taken literally,  (\ref{n0E}) and (\ref{Sigma0}) {\it cannot} be fulfilled simultaneously. 

We might simply proceed by fixing $M_0$ and $K$ each to a certain value consistent with the experimental data, and compute the surface tension for the 
resulting set of parameters. However, a single number for the surface tension in the given phenomenological model, and within the given approximations, would not be of particular 
significance since the relation to its actual value in QCD is unclear. 
Hence, we shall rather perform a more general study by investigating the phase structure and  
surface tension in the parameter space of the model. The idea is firstly to obtain 
a range of values for the surface tension, for instance asking for the maximal value that the model allows for, and secondly to search for 
qualitatively different scenarios. In doing so, we keep the range for $M_0$ and $K$ from 
Eq.\ (\ref{n0E}) in mind, but also include regions where $M_0$ and/or $K$ are beyond their "allowed" regime. While a given scenario might be outside the physical region in the given model, 
it may well occur in other phenomenological models, in more elaborate versions or approximations of the present model, or in QCD.

\begin{figure*} [t]
\begin{center}
\hbox{\includegraphics[width=0.5\textwidth]{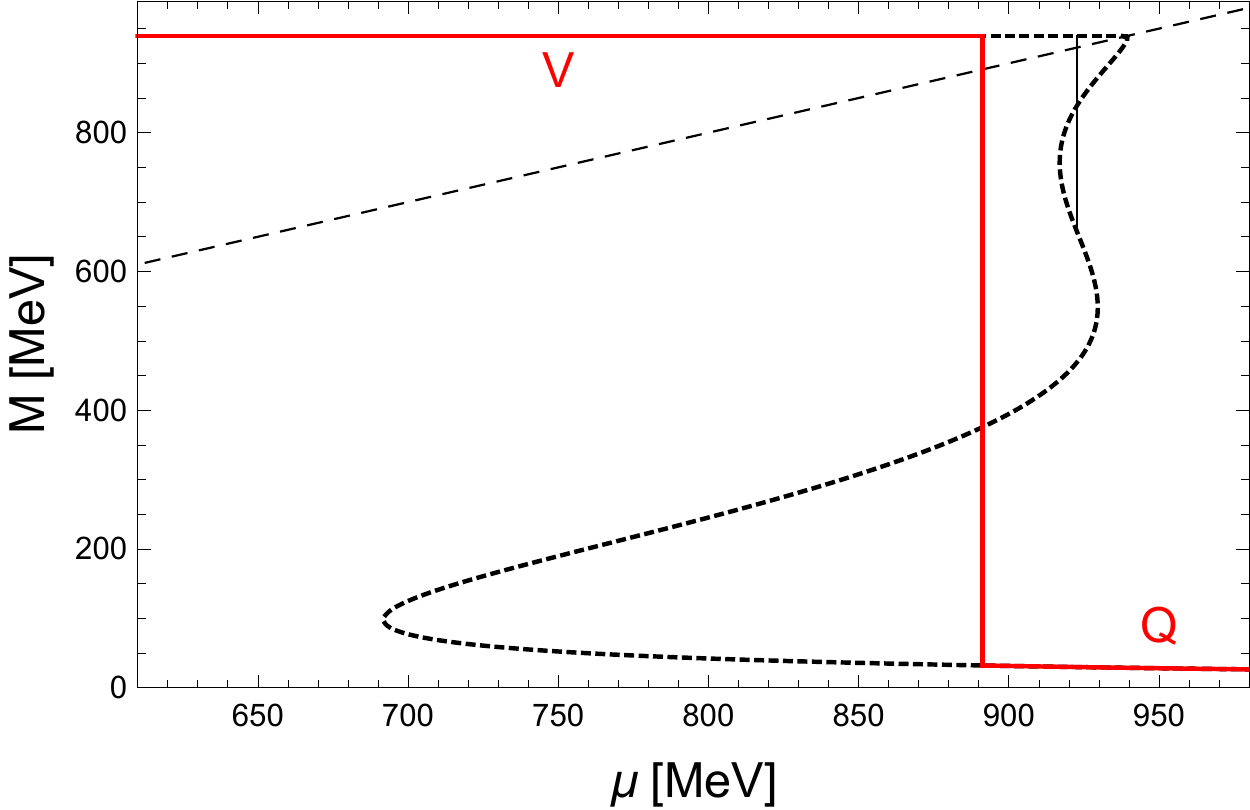}\includegraphics[width=0.5\textwidth]{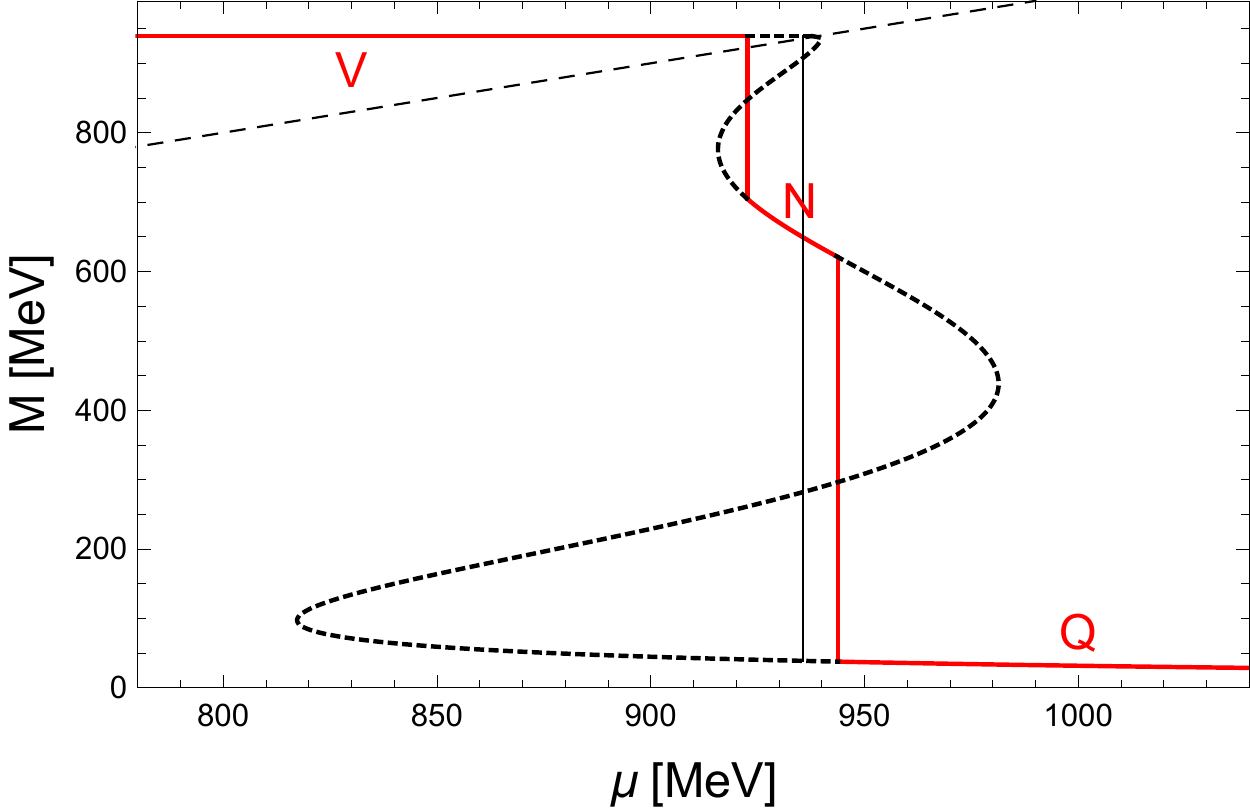}}
\hbox{\includegraphics[width=0.5\textwidth]{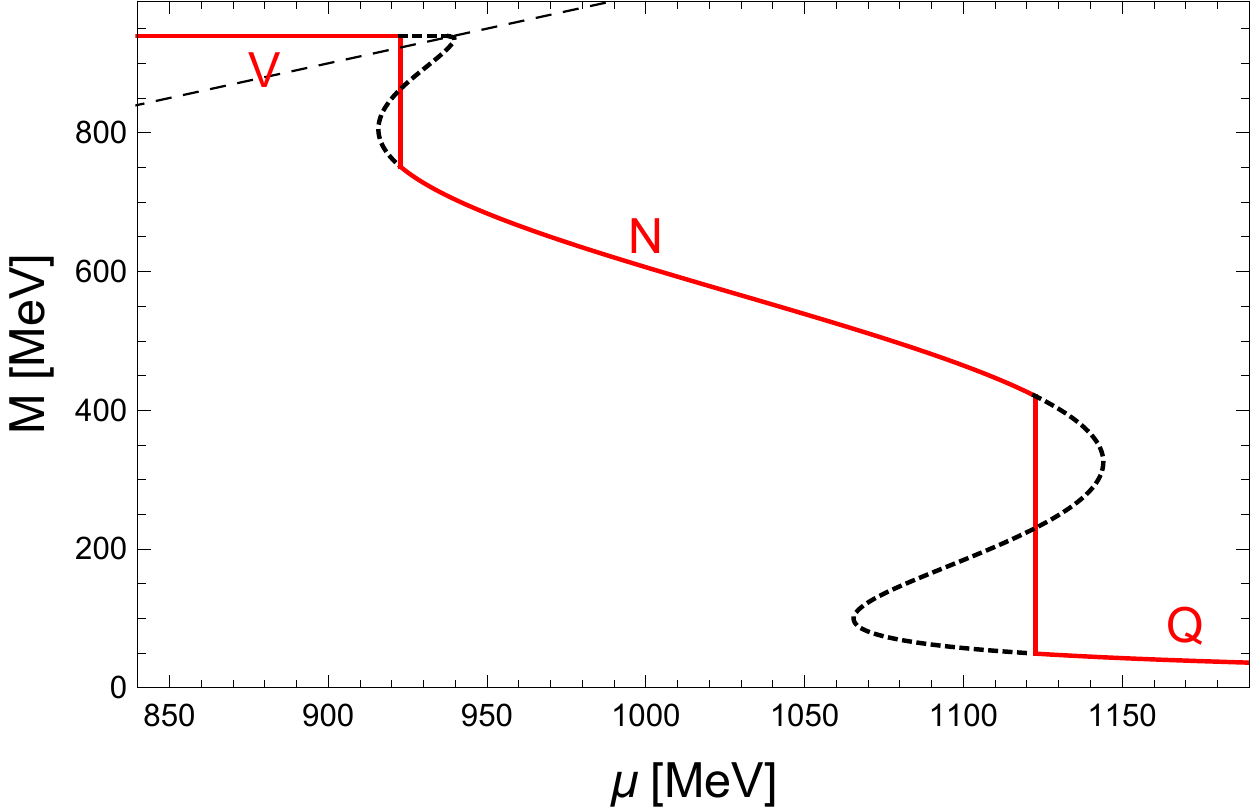}\includegraphics[width=0.5\textwidth]{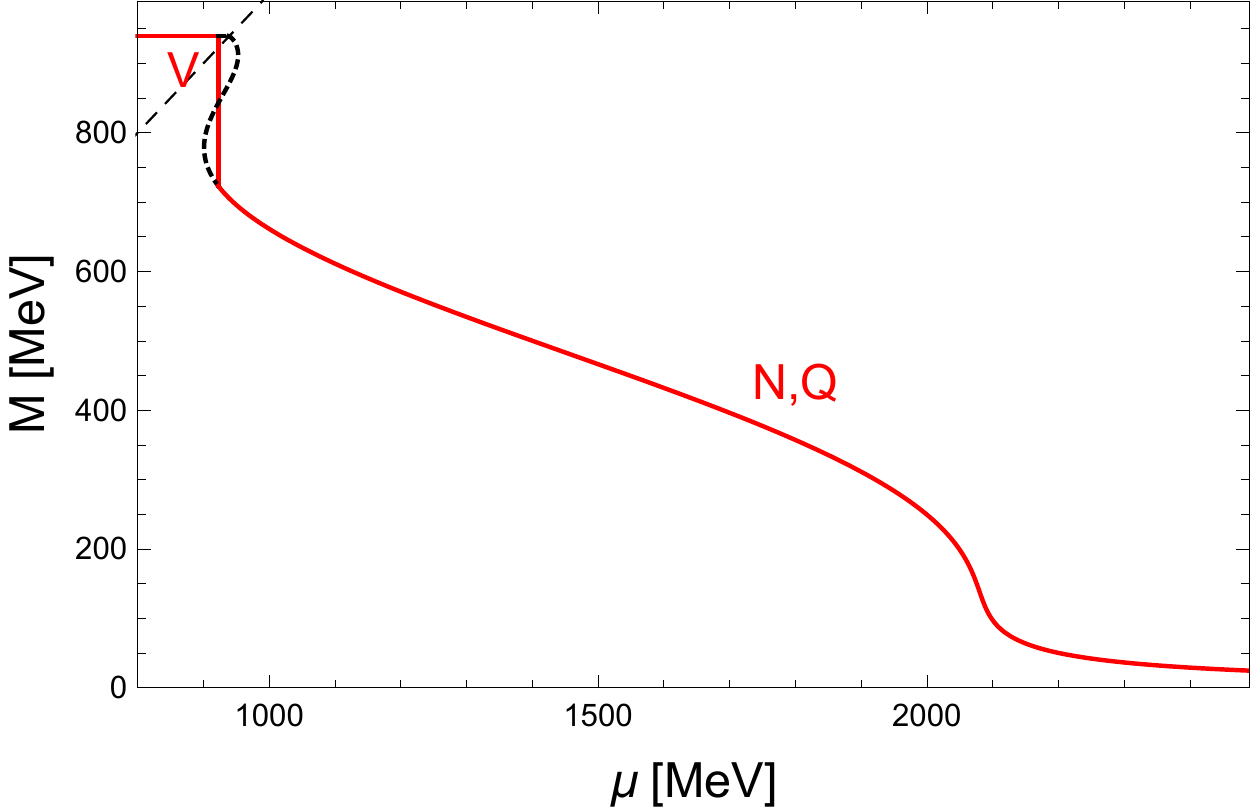}}
\caption{Effective nucleon mass as a (multivalued) function of the baryon chemical potential for zero temperature and four different parameter sets, showing three stable branches: V (vacuum), N (nuclear matter), and Q (chirally restored phase). The solid (red) lines indicate the preferred phase and show the first-order phase transitions, while the thick dashed lines correspond to metastable and unstable branches. The four panels show four qualitatively different cases: metastable nuclear matter (upper left panel), stable nuclear matter with a first-order NQ phase transition (upper right and lower left panel) and an NQ crossover (lower right panel); the qualitative difference between the upper right and the lower left panels is the (non-)overlap between the spinodal regions of the VN and NQ transitions. The parameters are $(M_0/m_N,K/{\rm MeV}) = (0.7, 200), (0.75, 300), (0.8, 300), (0.77, 760)$ in the order upper left, upper right, lower left, lower right. 
The thin dashed line indicates $\mu=M$, which separates the vacuum, where only the trivial 
solution $M=m_N$ exists, from the region with nontrivial solutions. In the upper panels there exists a chemical potential where there is phase coexistence between two phases which are not the global minimum, indicated by a thin (black) vertical line. 
}
\label{fig:Mmu}
\end{center}
\end{figure*}

\section{Homogeneous phases and phase transitions}
\label{sec:hom}

\begin{figure*} [t]
\begin{center}
\hbox{\includegraphics[width=0.5\textwidth]{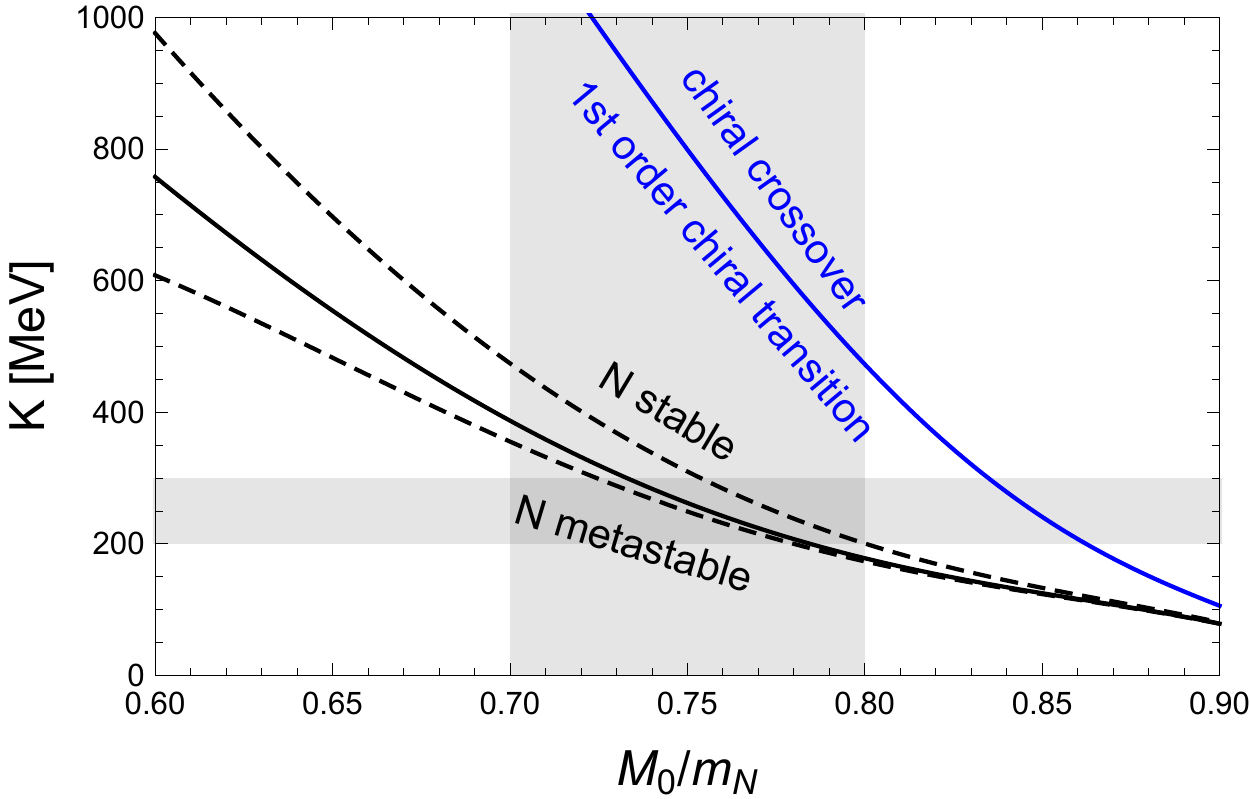}\includegraphics[width=0.5\textwidth]{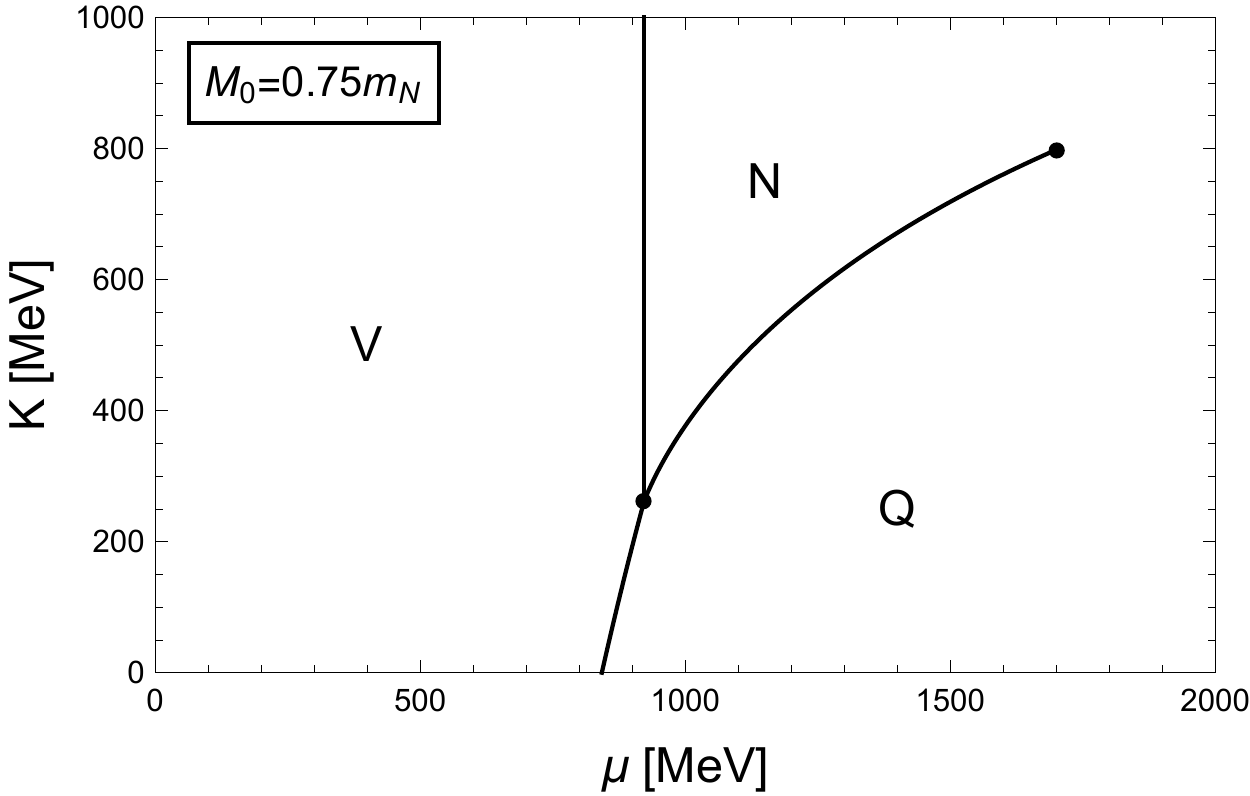}}
\caption{{\it Left panel:} Qualitatively different regimes at zero temperature in the plane of effective nucleon mass at saturation $M_0$ and incompressibility at saturation $K$. The lower (black) solid line separates the parameter region where nuclear matter is stable from the region where it is unstable. The upper (blue) solid line separates the regions where the transition between nuclear and quark matter is of first order and where it is a crossover. The dashed lines are relevant for the calculation of the surface tension at the first order chiral transition: above the upper dashed line there is no phase coexistence between the vacuum and the chirally restored phase for any $\mu$, and below the lower dashed line there is no phase coexistence between nuclear matter and the chirally restored phase for any $\mu$. The shaded bands indicate the allowed physical values for $M_0$ and $K$ according to Eq.\ (\ref{n0E}). {\it Right panel:} First order phase transition lines in the plane of incompressibility and chemical potential for a fixed $M_0$. Since the transition between nuclear matter 
and the vacuum is held fixed, the VN phase transition line is vertical. The topology of this diagram is the same for all $M_0$. In particular, for each $M_0$ there is a $K$ for which all three phase transitions occur at the same critical chemical potential, which defines the lower (black) solid line of the left panel.
}
\label{fig:paraspace}
\end{center}
\end{figure*}

As a preparation for the calculation of the surface tension, let us start by discussing the homogeneous phases. To that end, we solve the stationarity equations (\ref{dOm}) for a given baryon chemical potential $\mu$ and at temperature $T=0$ for constant $\bar{\sigma}$ and $\bar{\omega}$, so that $\nabla^2 \bar{\sigma}= \nabla^2 \bar{\omega}=0$. 

In Fig.\ \ref{fig:Mmu} we show the effective nucleon mass as a function of $\mu$, including all unstable and metastable branches, which arise due to the multivalued nature of this function\footnote{The parameter sets for the four panels of Fig.\ \ref{fig:Mmu}, corresponding to the given values for $M_0$ and $K$, are: $(g_\omega, a_2, a_3/{\rm MeV}^{-2}, a_4/{\rm MeV}^{-4}) = (10.60, 38.72, -8.084 \times 10^{-3}, 3.419 \times 10^{-5}), 
(9.467, 49.41, 4.064 \times 10^{-3}, 6.297 \times 10^{-5}), (8.161, 59.57, 1.343 \times 10^{-2},  1.159 \times 10^{-4}), (8.969, 123.5, 0.2469, 3.505 \times 10^{-4})$, in the order upper left, upper right, lower left, lower right.}.  
Recall that the nucleon mass is, up to a constant factor $g_\sigma$, the same as the sigma condensate. 
The omega condensate shows the same qualitative features that are relevant for the following discussion, and thus here we can restrict ourselves to showing the behavior of only one of the condensates.
 
We are mainly interested in the first-order phase transitions and the associated spinodal regions since this is where we shall compute the various surface tensions\footnote{The spinodal region of a first-order  transition between two phases A and B is defined as the range of chemical potentials 
where phase A exists as a metastable solution when phase B is the preferred phase and vice versa. Here, A and B can be either of the three potentially stable branches of the model, V, N, or Q.}. 
We abbreviate the 
three stable branches by V (vacuum), N (nuclear matter), and Q (chirally restored phase, "quark matter").
The baryon onset -- the transition between V and N -- is a first-order phase transition by construction. The chiral phase transition between N and Q can be of first order (upper right and lower left panels) or a crossover (lower right panel). This crossover only occurs for 
extremely large (unphysical) values of the incompressibility. 
Since our model breaks chiral symmetry explicitly, there are no second-order phase transitions in the model. 

For certain parameter choices there is a first-order transition between V and Q (upper left panel). Even if our model is, in principle, not suitable to account for realistic quark matter, since it contains no quark degrees of freedom and no strangeness, we can phenomenologically identify Q with the quark matter phase. Then, this scenario is a realization of the strange quark matter hypothesis, according to which QCD would feature a direct transition from the vacuum to strange quark matter, and nuclear matter would be metastable \cite{PhysRevD.4.1601,PhysRevD.30.272}. If we do not impose stability (as opposed to metastability) of nuclear matter as an additional constraint, the parameter regime where this hypothesis is realized has to be kept and included in the calculation of the surface tension. Surprisingly, as we shall see in Fig.\ \ref{fig:paraspace}, this regime overlaps significantly with the parameter regime allowed by the constraints (\ref{n0E}). 

As a result of this analysis, we can compute three different surface tensions: $\Sigma_{\rm NQ}$, $\Sigma_{\rm VQ}$, and $\Sigma_{\rm VN}$. We mainly do so at the phase transitions, but also discuss bubbles of the stable phase immersed in the metastable phase, for which the spinodal regions are relevant. 
Notice, in particular, that in the presence of two first-order phase transitions, VN and NQ, the spinodal regions can overlap (as in the upper right panel of Fig.\ \ref{fig:Mmu}). 

Fig.\ \ref{fig:paraspace} shows where the different scenarios of Fig.\ \ref{fig:Mmu} can be found in the $M_0$-$K$ plane. The left panel of this figure identifies the regions in which first-order NQ and VQ transitions exist,  
which is useful in the calculation of the surface tension. It is only below the upper dashed line, for instance, that there is phase coexistence between the V and Q phases. 
In other words, above this line there exists no chemical potential at which these phases  
have the same free energy. 
In the small regime between the upper dashed line and the lower (black) solid line, V and Q  have the same free energy at a chemical potential at which nuclear matter N is the ground state, and where the free energy minimum corresponding to the ground state N is between the minima corresponding to the metastable states V and Q. 
Therefore, in this regime, there is no stable domain wall configuration for the VQ transition and we shall not compute $\Sigma_{\rm VQ}$ there. 

Analogously, it is only above the lower dashed line that there exists, for each pair $(M_0,K)$, a chemical potential at which N and Q coexist. Again, there is a small stripe in the $M_0$-$K$ plane where a third phase -- here the vacuum V -- is preferred at the point where N and Q have the same free energy. In this case, however, the ground state has an effective nucleon mass that is {\it not} in between the effective nucleon masses of the two metastable states, and thus we do find stable domain wall configurations. 

Finally, we see 
that the surface tension of the NQ transition must approach zero for sufficiently large $K$ and/or $M_0$, since the first-order 
phase transition line turns into a crossover in that region.  
The right panel of Fig.\ \ref{fig:paraspace} provides 
a complementary view of the different phases in the plane of incompressibility versus chemical potential. 

In summary, the surface tensions exactly at the critical chemical potentials can be calculated in the following regimes: $\Sigma_{\rm NQ}$ can be calculated anywhere between the lower dashed line and the upper (blue) solid line 
and $\Sigma_{\rm VQ}$ can be calculated anywhere below the lower (black) solid line. Since there is always a first-order transition between V and N by construction, we can compute $\Sigma_{\rm VN}$ throughout the parameter space, although this includes regions where V and N are metastable phases. As mentioned above, the "allowed" regime, given by the shaded rectangle where the two shaded bands intersect, contains a significant fraction (about half of its area), where nuclear matter does not exist as a stable phase.

\section{Domain walls, bubbles and the surface tension}
\label{sec:calc}

The surface tension is obtained from classical configurations connecting different homogeneous phases, which satisfy Eqs. (\ref{dOm}) \cite{doi:10.1063/1.1730447,Langer:1969bc,LANGER197461,domb1983phase}. It can be calculated in different geometries and using different methods. 
The geometries we consider here are domain walls and bubbles. For the former, the surface tension is unambiguously defined as the free energy difference per unit area between the domain wall configuration and the homogeneous configuration of either phase, i.e., the surface tension is the energy cost that arises from abandoning the first minimum and walking uphill and downhill through the potential to reach the second minimum. On the other hand, stable solutions for a bubble are obtained in the spinodal region, where both phases exist as local minima but have different free energies. In this case, the condensates interpolate between the "false vacuum" (far away from the bubble at $r=\infty$) and some value close to the "true vacuum" (in the center of the  bubble at $r=0$). As we approach the phase transition, the value inside the bubble approaches the true vacuum and the radius of the bubble, which has to be determined dynamically, approaches infinity, i.e., we approach the domain wall solution \cite{Coleman:1977py}.

In the present section, we select specific parameter sets to present the domain wall and bubble profiles, discussing nonzero-temperature effects, approximations and their relation to the full numerical result. These approximations are addressed in some detail due to the existence of two condensates in the present model. If only a single condensate were present the calculation would be straightforward. Numerical calculations of the domain wall or bubble profiles in related models with more than one condensate can be found in Refs.\ \cite{Menezes:1999zz,PhysRevD.84.036011}. 

\subsection{Domain walls}

In the domain wall geometry the Euler-Lagrange equations (\ref{dOm}) become 
\begin{subequations}\label{dOmD}
\bea
\frac{d^2\bar{\sigma}}{dx^2} &=& \frac{\partial U}{\partial\bar{\sigma}} + g_\sigma n_s \, ,  \label{eom1D} \\[2ex] 
\frac{d^2\bar{\omega}}{dx^2} &=& 
 m_\omega^2\bar{\omega} - g_\omega n_B  \, . \label{eom2D}
\eea
\end{subequations} 
After having identified the first-order phase transitions in Sec.\ \ref{sec:hom}, we can solve this system of differential equations at the phase transition with the boundary conditions $\bar{\sigma}(x=\pm\infty) = \bar{\sigma}_\pm$, $\bar{\omega}(x=\pm\infty) = \bar{\omega}_\pm$, where the pairs $(\bar{\sigma}_-,\bar{\omega}_-)$ and $(\bar{\sigma}_+,\bar{\omega}_+)$ are the solutions of the homogeneous equations, corresponding to the two phases that have the same free energy at the phase transition \cite{Coleman:1977py}. We solve the equations numerically via successive over-relaxation (a useful method employed in similar contexts, e.g., in the calculation of flux tube profiles \cite{Haber:2017kth,Haber:2017oqb}).

\begin{figure*} [t]
\begin{center}
\hbox{\includegraphics[width=0.5\textwidth]{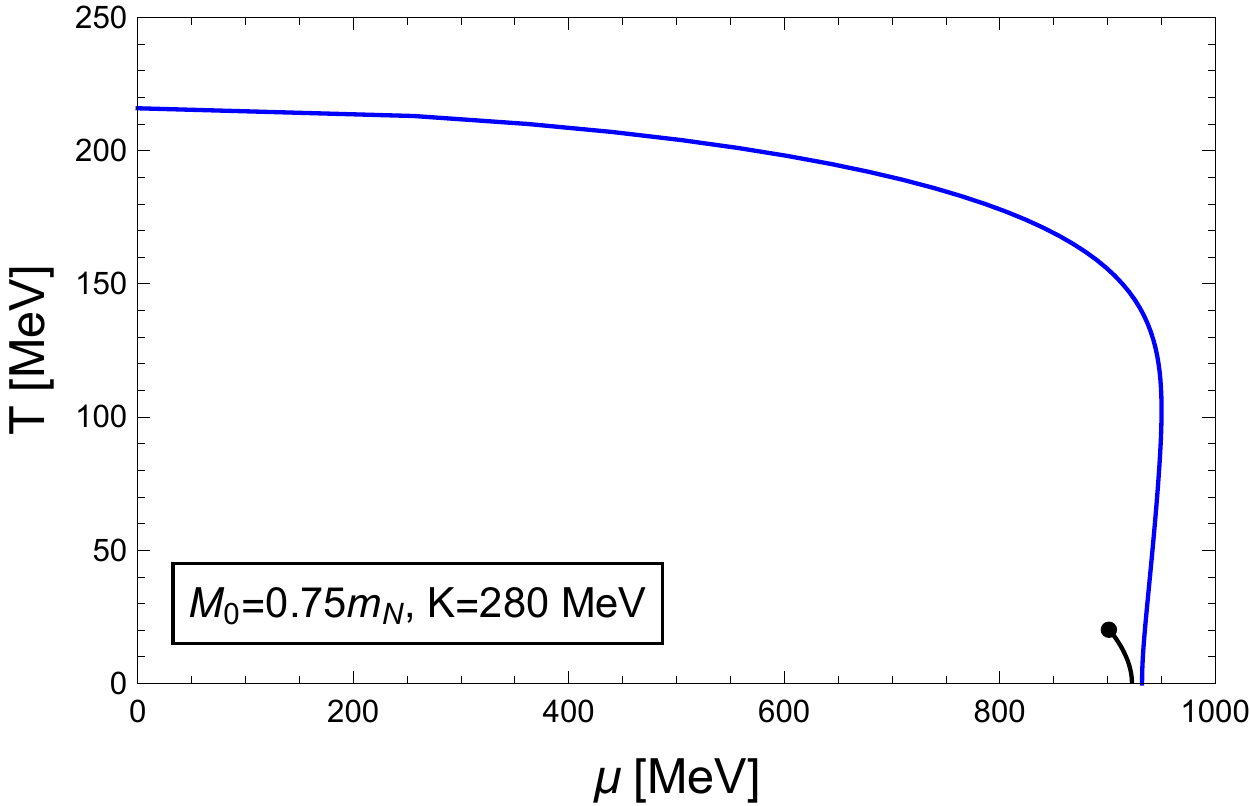}\includegraphics[width=0.5\textwidth]{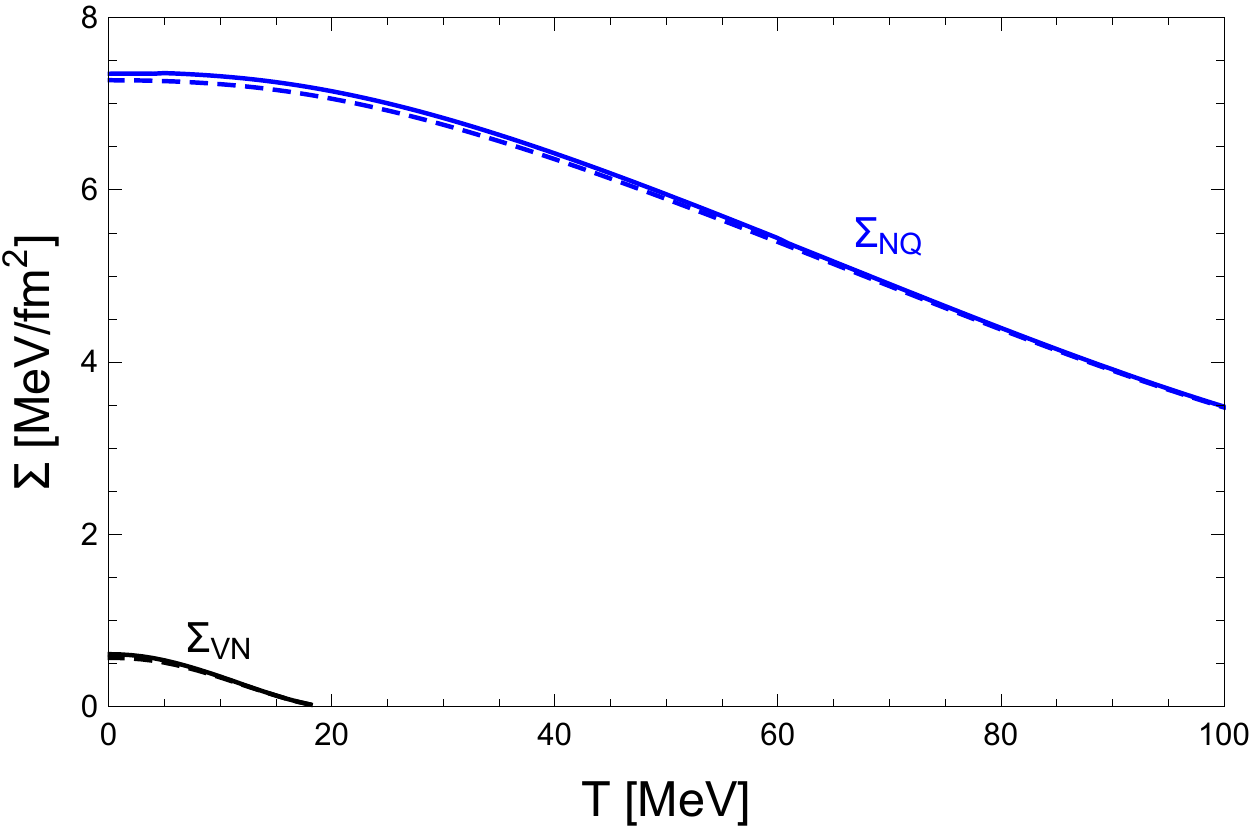}}
\hbox{\includegraphics[width=0.5\textwidth]{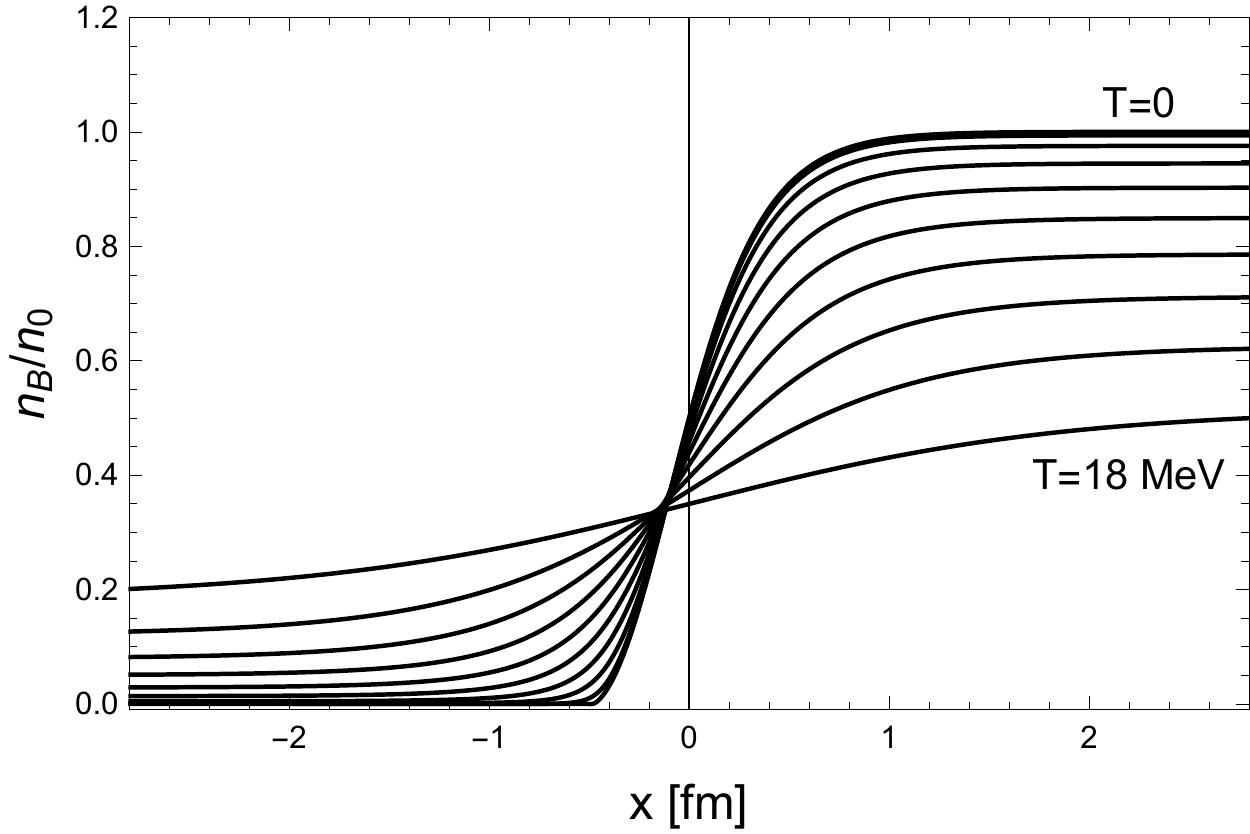}\includegraphics[width=0.5\textwidth]{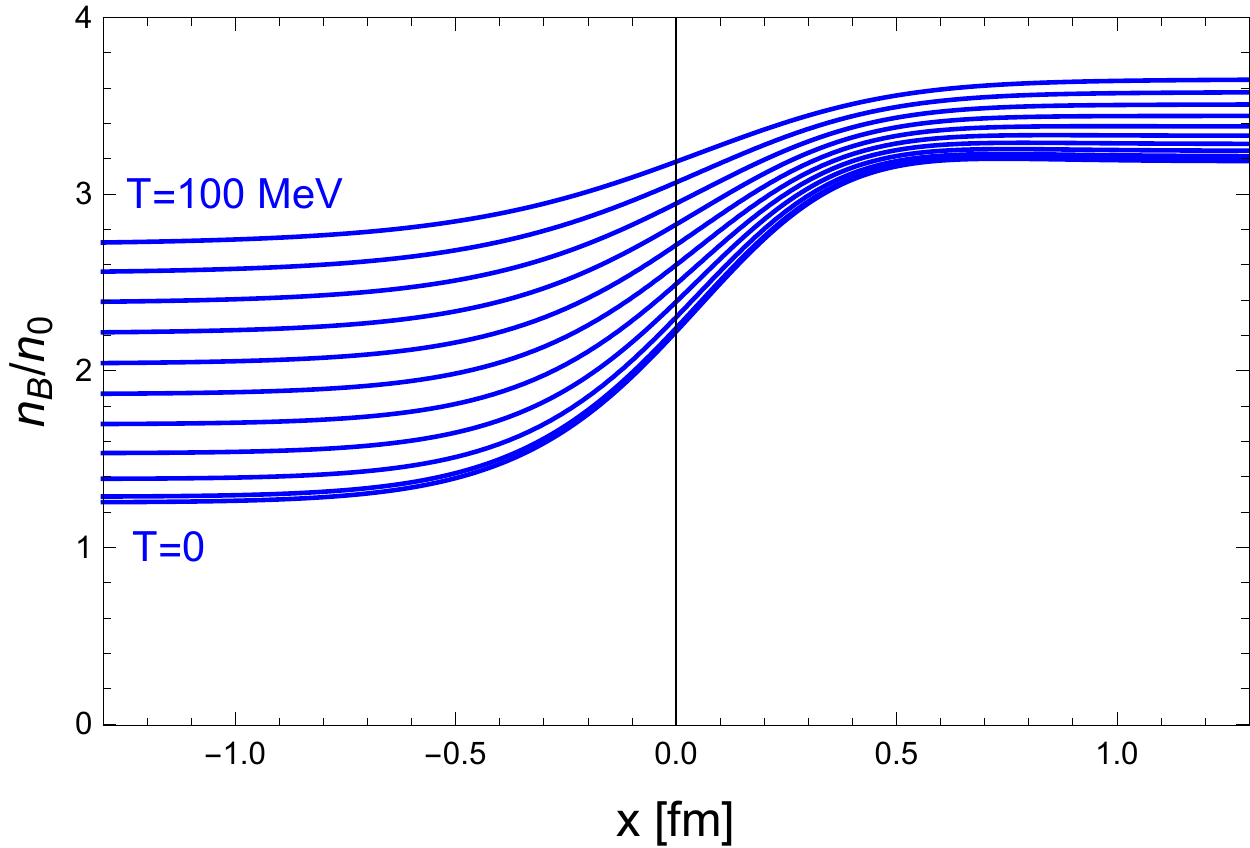}}
\caption{ {\it Upper left panel:} 
Chiral and liquid-gas phase transitions in the $\mu$-$T$-plane. {\it Upper right panel:} 
Surface tensions as a function of temperature along the phase transition lines
for the liquid-gas (vacuum-nuclear, VN) and chiral (nuclear-quark, NQ) transitions. 
{\it Lower panels:} Numerically computed profiles of the 
baryon density $n_B$ in units of the saturation density $n_0$ across the domain wall for different temperatures 
for the liquid-gas phase transition (left panel, in temperature steps of 2 MeV) and the chiral transition (right panel, in temperature steps of 10 MeV). In all panels, $M_0=0.75 m_N$, $K=280\, {\rm MeV}$.}
\label{fig:profiles}
\end{center}
\end{figure*}

Once the numerical solution is obtained, the surface tension is computed from 
\bea \label{Sigma}
 \Sigma  &=& \int_{-\infty}^\infty dx \Bigg[\frac{1}{2}\left(\frac{d\bar{\sigma}}{dx}\right)^2 -\frac{1}{2}\left(\frac{d\bar{\omega}}{dx}\right)^2 \non [2ex] 
 && \hspace{1.2cm} +\,\Omega_0(\bar{\sigma},\bar{\omega})-\Omega_{\rm hom} \Bigg] \, , 
 \eea
where $\Omega_{\rm hom}\equiv \Omega_0(\bar{\sigma}_-,\bar{\omega}_-) =\Omega_0(\bar{\sigma}_+,\bar{\omega}_+)$ is the free energy of either  homogeneous phase 
far away from the domain wall. The form of the surface tension (\ref{Sigma}) follows immediately 
from the free energy density (\ref{Omega}) because $\Sigma$ is defined as the free energy difference per unit area between the domain wall configuration and the 
homogeneous phase. The brute force numerical calculation is the most direct way to compute the surface tension; 
within the given Thomas-Fermi, mean-field, and no-sea approximations (and up to negligibly small numerical inaccuracies), it corresponds to the exact result.

We show the domain wall profiles for a certain choice of the parameters $M_0$ and $K$ and various temperatures in Fig.\ \ref{fig:profiles}. The parameter choice is such that nuclear matter is stable and at zero temperature the baryon onset is succeeded by a first order chiral transition. 

In the upper left panel we extend these two first-order phase transitions into the $T$-$\mu$ plane to show the critical chemical potentials at which the surface tension is calculated. We  plot the entire chiral phase transition line, up to its intersection with the temperature axis, where it remains of first order, in contradiction with lattice results for QCD. One can check that, within our approximation, this is the case for the entire parameter space (in contrast to the chemical potential axis, where the chiral transition {\it can} become a crossover, as discussed above). However, this apparent contradiction with QCD is irrelevant since we should not trust our mean-field calculation for large temperatures. In what follows, we restrict ourselves to temperatures below $100$ MeV, having in mind that even at these temperatures fluctuations might yield significant corrections to our result. 

In the upper right panel the surface tension is shown as a function of temperature. Not surprisingly, it decreases with increasing temperature. In particular, it vanishes in the case of the VN transition\footnote{At nonzero temperature, the terminology V ("vacuum") becomes inappropriate, nevertheless we have kept this notation for simplicity, V being the stable branch that is continuously connected to the zero-temperature vacuum branch.} at the critical point which, for the given parameters, occurs at around $T\simeq 19.6\, {\rm MeV}$, in good agreement with experiment and more sophisticated nuclear physics calculations (see, e.g., Ref.\ \cite{Carbone:2018kji} and references therein). Notice that our zero-temperature surface tension $\Sigma_{\rm VN}$ is smaller than that of real-world nuclear matter (\ref{Sigma0}) for the given parameter set. We shall see below that one can only fulfill Eq.\ (\ref{Sigma0}) in a parameter regime where nuclear matter is metastable or by going beyond the regime given by Eq.\ (\ref{n0E}). The dashed lines in the upper right panel, barely distinguishable from the solid lines, 
show the result of an approximation which we explain now.

\subsection{Approximations for the surface tension}

\subsubsection{Semi-analytical approximation}

It is useful to implement a simpler approximation for the surface tension, which does not require a numerical solution of the coupled differential equations. As Fig.\ \ref{fig:profiles} suggests, the results of the following approximation are very close to the full numerical results. And, in any case, we should keep in mind that we have already employed various approximations to set up the
profile equations, which thus are not exact to begin with.

The approximation is derived with the help of the first integral of motion
\be \label{sw}
\frac{1}{2}\left(\frac{d\bar{\sigma}}{dx}\right)^2 -\frac{1}{2}\left(\frac{d\bar{\omega}}{dx}\right)^2 - \Omega_0(\bar{\sigma},\bar{\omega}) = -\Omega_{\rm hom} 
\, , 
\ee
so that we can write the surface tension (\ref{Sigma}) as 
\bea \label{Sigma1}
\Sigma  &=& \int_{-\infty}^\infty dx \, \left[\left(\frac{d\bar{\sigma}}{dx}\right)^2 -\left(\frac{d\bar{\omega}}{dx}\right)^2\right] \non[2ex]
&=&\int_{-\infty}^\infty dx \, \left(\frac{d\bar{\sigma}}{dx}\right)^2(1- \bar{\omega}'^2)
 \non[2ex]
&=& \int_{\bar{\sigma}_-}^{\bar{\sigma}_+} d\bar{\sigma} \, \sqrt{2(\Omega_0 - \Omega_{\rm hom}) (1- \bar{\omega}'^2)}  \, . 
\eea
Assuming $\bar{\sigma}$ and $\bar{\omega}$ to be monotonic functions of $x$, we have introduced the
function $\bar{\omega}(\bar{\sigma})$, denoted its derivative with respect to $\bar{\sigma}$ by a prime, and, in the last step,  rewritten the spatial derivative of $\bar{\sigma}$ with the help of Eq.\ (\ref{sw}).

So far, this is merely an alternative way of writing the surface tension, and not much is gained yet because 
the function $\bar{\omega}(\bar{\sigma})$, which appears in $\Omega_0$ and whose derivative appears explicitly, can only be obtained from the full solution of the differential equations. Notice that this is different in the case of a single condensate, where Eq.\ (\ref{Sigma1}) only requires knowledge of the potential itself and reduces the calculation of the surface tension to a numerical integration. 

For two condensates, we can simplify the calculation of the surface tension by approximating $\bar{\omega}(\bar{\sigma})$: we drop the gradient of $\bar{\omega}$ in Eq.\ (\ref{eom2D}), such that Eq.\ (\ref{eom2D}) becomes $m_\omega^2 \bar{\omega} = g_\omega n_B$, which can be solved to find $\bar{\omega}(\bar{\sigma})$. This has to be done numerically, even at zero temperature, where the equation becomes a sixth-order polynomial in $\bar{\omega}$. By taking the derivative with respect to $\bar{\sigma}$ on both sides of this equation, we obtain a semi-analytical expression for the derivative [containing the numerical function 
$\bar{\omega}(\bar{\sigma})$],
\bea
\bar{\omega}'_{\rm app} &=& g_\omega \frac{\partial n_B}{\partial \bar{\sigma}}\left(m_\omega^2-g_\omega \frac{\partial n_B}{\partial \bar{\omega}}\right)^{-1}  \non[2ex]
&&\stackrel{T=0}{\longrightarrow}\;\;  -\frac{2g_\omega g_\sigma Mk_F\Theta(\mu_*-M)}{\pi^2 m_\omega^2+2g_\omega^2\mu_* k_F} \, .
\eea
Here we have added the subscript "app" to emphasize that this expression is approximate because it is {\it not} identical to the derivative one obtains from the full numerically calculated domain wall profiles. We can thus write our approximate result as 
\bea \label{SigmaApp}
\Sigma  
&\simeq& \int_{\bar{\sigma}_-}^{\bar{\sigma}_+} d\bar{\sigma} \, \sqrt{2(\Omega_0 - \Omega_{\rm hom}) (1- \bar{\omega}'^2_{\rm app})}  \, . 
\eea
We refer to this approximation as the "semi-analytical approximation". It is used  for the dashed lines in the upper right panel of Fig.\ \ref{fig:profiles}, which are in excellent agreement with the full result, and we shall also make use of it in Sec.\ \ref{sec:results}.

\begin{table*}[t]
\begin{center}
\begin{tabular}{|c||c|c|} 
\hline
\rule[-1.5ex]{0em}{5ex} 
  & domain walls & bubbles  \\[1ex] \hline\hline
\rule[-1.5ex]{0em}{6ex} 
full two-condensate solution & relax, Eq.\ (\ref{Sigma}) & -- \\[2ex] \hline
\rule[-1.5ex]{0em}{6ex} 
$\;\;$ semi-analytical approximation $\;\;$& $\;\;$ integrate, Eq.\ (\ref{SigmaApp}) $\;\;$ & -- 
\\[2ex] \hline
\rule[-1.5ex]{0em}{6ex} 
$\;\;$ one-condensate approximation $\;\;$  & $\;\;$ integrate, Eq.\ (\ref{Sigma2}) $\;\;$ & $\;\;$
shoot, Eq.\ (\ref{SigmaBubble}) $\;\;$ \\[2ex] \hline
\end{tabular}
\caption{Overview of methods and approximations used to calculate the surface tension $\Sigma$
in the planar (domain wall) and spherical (bubble) geometry, together with the  equations that show the respective expressions for $\Sigma$. The first two lines  are used for our main results in Sec.\ \ref{sec:results} and shown to be in very good agreement
with each other.
The approximation of the third line is only used in Fig.\ \ref{fig:bubbles}, since the methods from first and second lines fail for bubbles. "Relax" and "shoot" refer to full numerical 
calculations of the profiles with relaxation and shooting algorithms. 
"Integrate" refers to a numerical integration, without having to solve any differential equation. 
 }
\label{table:methods}
\end{center}
\end{table*}

\subsubsection{One-condensate approximation}

Another useful approximation is to replace Eq.\ (\ref{eom2D}) with $m_\omega^2 \bar{\omega} = g_\omega n_B$ from the beginning and use the resulting function $\bar{\omega}(\bar{\sigma})$ in the 
Euler-Lagrange equation for $\bar{\sigma}$ (\ref{eom1D}).
We can then compute the domain wall profile by solving a single differential equation numerically or, equivalently, by rederiving an expression for the surface tension that only requires a numerical integral. Were we only interested in the domain wall geometry, this would not be necessary since, as we have just demonstrated, we can solve the full system numerically with the relaxation method or can employ the semi-analytical approximation (\ref{SigmaApp}). However, we were not able to make the relaxation method work for the 
coupled differential equations in the case of the bubble geometry, where it turned out to be difficult to prevent the iterative procedure to relax to the trivial solution. Therefore, we shall make use of this 
"one-condensate approximation" in the 
calculation of the bubble profiles. Nevertheless, let us first derive the surface tension within this approximation in the domain wall geometry, as a foundation to define the surface tension of bubbles below. The difference with respect to the previous derivation is that now Eq.\ (\ref{sw}) becomes 
\be \label{sw1}
\frac{1}{2}\left(\frac{d\bar{\sigma}}{dx}\right)^2 - \Omega_0(\bar{\sigma},\bar{\omega}(\bar{\sigma})) = -\Omega_{\rm hom} 
\, .
\ee
Inserting this into the surface tension (\ref{Sigma}) yields 
\bea \label{Sigma2}
\Sigma  &=& \int_{-\infty}^\infty dx \, \left[\left(\frac{d\bar{\sigma}}{dx}\right)^2 -\frac{1}{2}\left(\frac{d\bar{\omega}}{dx}\right)^2\right] \non[2ex]
&=& \int_{\bar{\sigma}_-}^{\bar{\sigma}_+} d\bar{\sigma} \, \sqrt{2(\Omega_0 - \Omega_{\rm hom})} \left(1- \frac{\bar{\omega}'^2_{\rm app}}{2}\right) \, . 
\eea
Notice that once we have replaced the Euler-Lagrange equation for $\bar{\omega}$, Eqs.\ (\ref{sw1}) and (\ref{Sigma2}) follow without further approximations. In particular, now there is only one function $\bar{\omega}'=\bar{\omega}'_{\rm app}$ since we do not solve the two-condensate differential equations.

\subsection{Bubbles}

Assuming spherical symmetry, Eqs.\ (\ref{dOm}) become 
\begin{subequations}\label{dOmB}
\bea
\frac{d^2\bar{\sigma}}{d r^2} +\frac{2}{r}\frac{d\bar{\sigma}}{dr} &=& \frac{\partial U}{\partial\bar{\sigma}} + g_\sigma n_s \, ,  \label{eom1B} \\[2ex] 
\frac{d^2\bar{\omega}}{d r^2} +\frac{2}{r}\frac{d\bar{\omega}}{dr} &=& 
 m_\omega^2\bar{\omega} - g_\omega n_B  \, . \label{eom2B}
\eea
\end{subequations}  
To compute the profiles of stable bubbles, one first needs to identify the spinodal region around the first-order phase transition. Then, a stable bubble profile is found by solving Eqs.\ (\ref{dOmB}) with the boundary conditions $\bar{\sigma}(\infty) = \bar{\sigma}_\infty$, $\bar{\omega}(\infty) = \bar{\omega}_\infty$, $\left.\frac{d\bar{\sigma}}{dr}\right|_{r=0} = \left.\frac{d\bar{\omega}}{dr}\right|_{r=0}=0$, where $(\bar{\sigma}_\infty,\bar{\omega}_\infty)$ is the solution of the homogeneous equations corresponding to the phase with the {\it larger} free energy ("false vacuum"). The values 
of the condensates in the center of the bubble have to be determined dynamically \cite{Coleman:1977py}.

In the calculation of the bubble profiles we employ the one-condensate approximation discussed above. We solve the differential equation for $\bar{\sigma}$ straightforwardly with a shooting algorithm. The surface tension of the bubble is not uniquely defined because there is no homogeneous reference state, unless one defines a sharp bubble radius. We  require the surface tension of a bubble to approach the surface tension of the domain wall at the phase transition. In the given one-condensate approximation, this surface tension is given by Eq.\ (\ref{Sigma2}), so that we may define the bubble surface tension as
\be \label{SigmaBubble}
\Sigma  = \int_0^\infty dr\,\left[\left(\frac{d\bar{\sigma}}{dr}\right)^2 -\frac{1}{2} \left(\frac{d\bar{\omega}}{dr}\right)^2\right] \, ,
\ee
where $\bar{\omega}(r) = \bar{\omega}(\bar{\sigma}(r))$ with $\bar{\omega}(\bar{\sigma})$ from 
$m_\omega^2 \bar{\omega} = g_\omega n_B$ and $\bar{\sigma}(r)$ determined numerically from solving for the bubble profile. One might argue that, 
even though we have approximated the problem by a single differential equation, we could as well use the original expression (\ref{Sigma1}) for the surface tension. However, we have found that rederiving the surface tension consistently within the one-condensate approximation gives a much better approximation to the full result. To keep track of the various methods and approximations used in this paper, we have collected them in Table \ref{table:methods}. 

\begin{figure*} [t]
\begin{center}
\hbox{\includegraphics[width=0.5\textwidth]{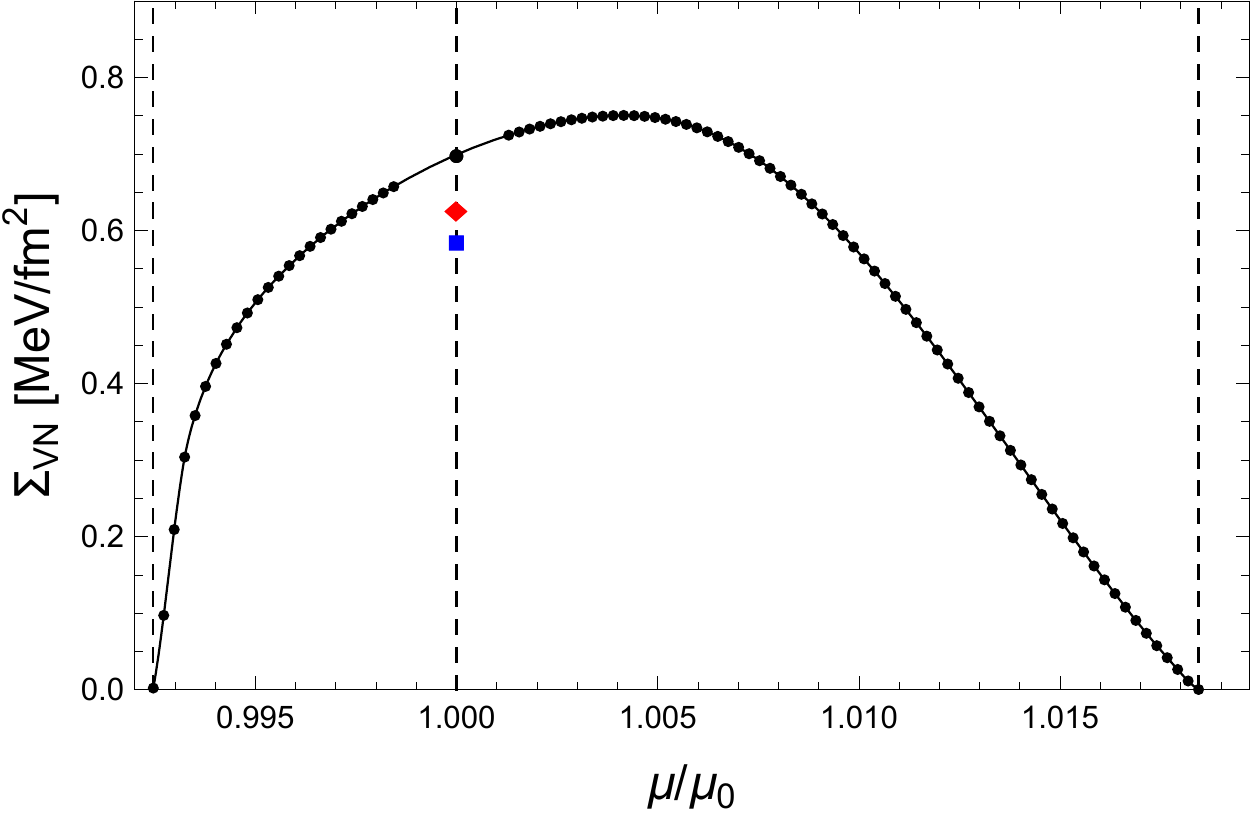}\includegraphics[width=0.5\textwidth]{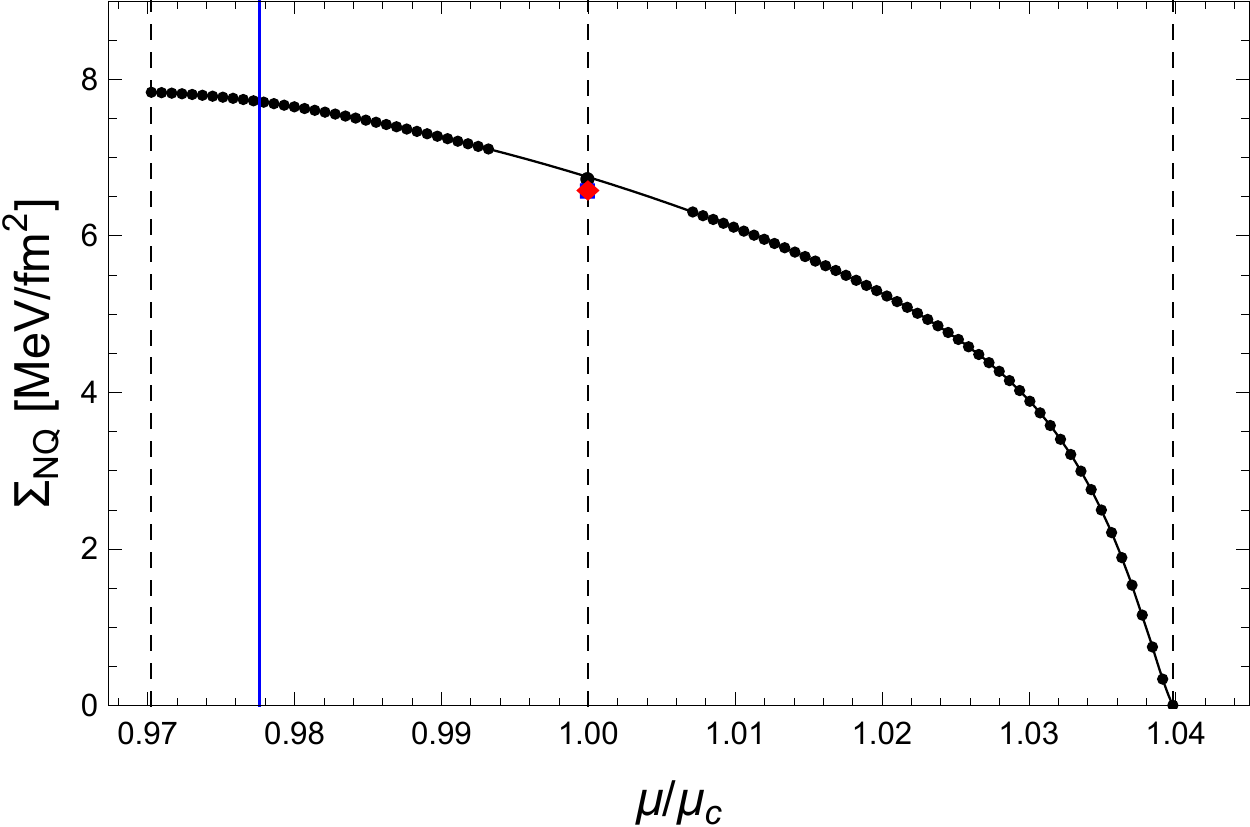}}
\caption{Zero-temperature surface tension of bubbles in the spinodal region of the VN (left panel) and NQ (right panel) transitions for $M_0=0.75\,  m_N$, $K=300\, {\rm MeV}$. The (black) spheres in the spinodal region (bounded by the outer dashed lines) are the results from the numerical calculation of the bubble profiles within the one-condensate approximation, the (red) diamonds are the results from the full numerical domain wall calculation, and the (blue) squares are from the semi-analytical approximation (\ref{SigmaApp}). The additional (blue) vertical line in the right panel indicates the baryon onset, i.e., to the left of the line nuclear matter is metastable. The critical chemical potentials are $\mu_0=922.7\, {\rm MeV}$ (left) and $\mu_c=943.8\, {\rm MeV}$ (right). }
\label{fig:bubbles}
\end{center}
\end{figure*}

In Fig.\ \ref{fig:bubbles} we plot the surface tension for the spinodal regions of the VN (left panel) and NQ (right panel) transitions for $M_0=0.75 m_N$ and $K=300\, {\rm MeV}$. These parameters are the same as in the upper right panel of Fig.\ \ref{fig:Mmu}. In the left panel, for chemical potentials smaller than the onset chemical potential $\mu_0$, stable vacuum bubbles exist in metastable nuclear matter N, while for chemical potentials larger than $\mu_0$ nuclear matter bubbles exist in the metastable vacuum V. In the right panel, analogously, we have N bubbles in Q below $\mu_c$ and Q bubbles in N above $\mu_c$. The (black) dots away from the phase transition are the results from the numerical calculation of the bubble profiles. 

As the phase transition is approached, the numerics get more challenging because the bubbles become larger and the change in the condensate occurs in a very small range of $r$ compared to the range that needs to be considered in the calculation. Therefore, there are no results close to the phase transition, and we have employed an interpolation to cover this regime (thin solid line). Exactly at the phase transition we can calculate the surface tension from the domain wall profile, either by solving the single differential equation in the planar geometry or, equivalently, by using the second line of Eq.\ (\ref{Sigma2}). This result is also shown as a (black) dot. Its agreement with the interpolated result is a check for the shooting algorithm. We have also indicated the result at the phase transition from the full two-condensate calculation [(red) diamond] and from the semi-analytical approximation (\ref{SigmaApp}) [(blue) square]. We see that in the case of the chiral phase transition all results are in very good agreement, while for the baryon onset there is a  deviation of the order of 10\%\footnote{In Ref.\ \cite{Drews:2013hha}, the surface tension $\Sigma_{\rm VN} \simeq 1.1\, {\rm MeV}\, {\rm fm}^{-2}$ is quoted for a very similar parameter set as we use for Fig.\ \ref{fig:bubbles}. Our result (say from the full domain wall calculation, but also from the approximations) is significantly smaller compared to that value, almost by a factor 1/2. We found that if we work with the approximation of calculating $\bar{\omega}(\bar{\sigma})$ from the homogeneous equation $m_\omega^2 \bar{\omega} = g_\omega n_B$, but then completely ignore the $\bar{\omega}'$ term in the surface tension, i.e., if we use
\bea \label{approx2}
\Sigma &\simeq & \int_{\bar{\sigma}_-}^{\bar{\sigma}_+} d\bar{\sigma} \, \sqrt{2(\Omega_0 - \Omega_{\rm hom})}  \, , \nonumber
\eea
we do obtain $\Sigma_{\rm VN} \simeq 1.07\, {\rm MeV}\, {\rm fm}^{-2}$, in agreement with Ref.\ \cite{Drews:2013hha}. It is clear from the comparison with the full result that this approximation is too simplistic.}. 

There is an obvious qualitative difference between the left and right panels of Fig.\ \ref{fig:bubbles} regarding the behavior at the boundaries of the spinodal region. The "standard" scenario is that the spinodal region is bounded from both sides by points at which the second, metastable solution ceases to exist. Therefore, as we move to the edges of the spinodal region, the local minimum that is assumed far away from the bubble becomes more and more shallow. As a consequence, the bubble profile flattens out and as we approach the spinodal boundary the surface tension approaches zero. This scenario is borne out in the left panel. 

\begin{figure*} [t]
\begin{center}
\hbox{\includegraphics[width=0.5\textwidth]{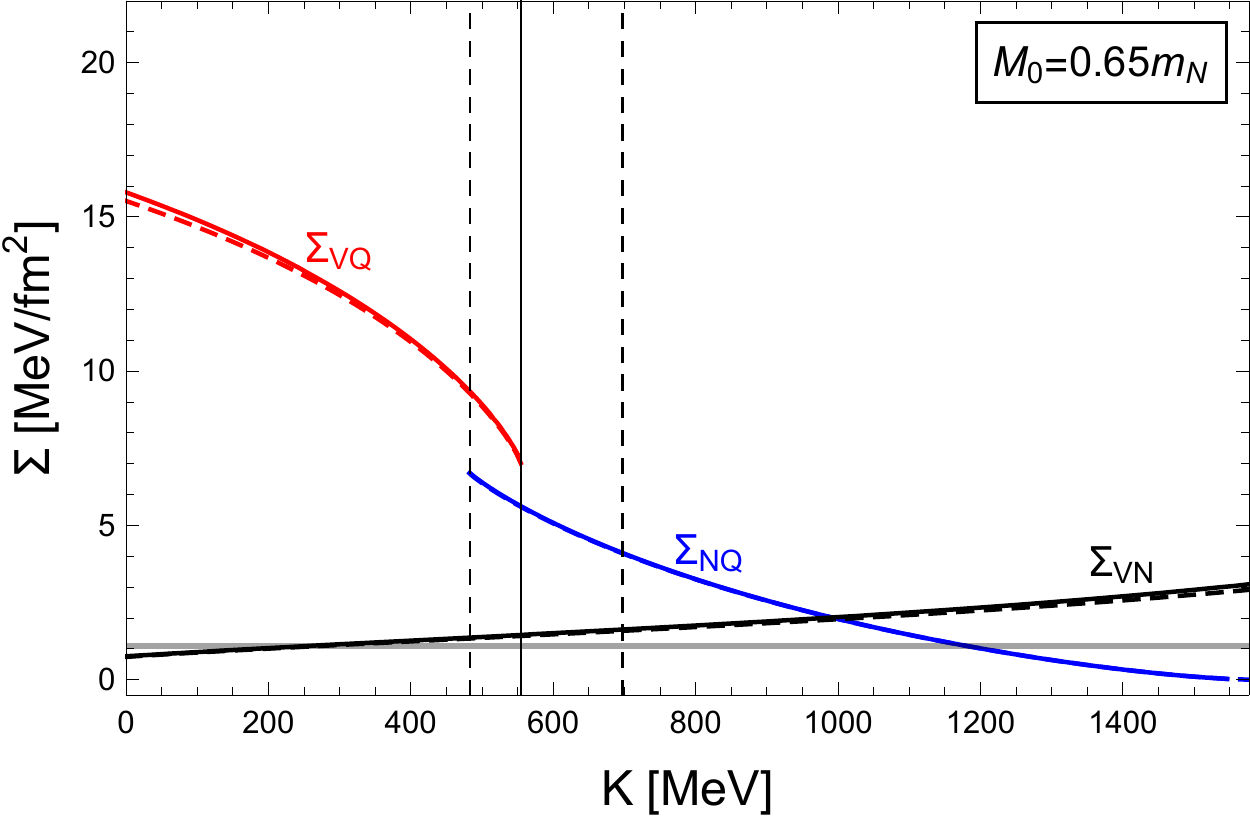}\includegraphics[width=0.5\textwidth]{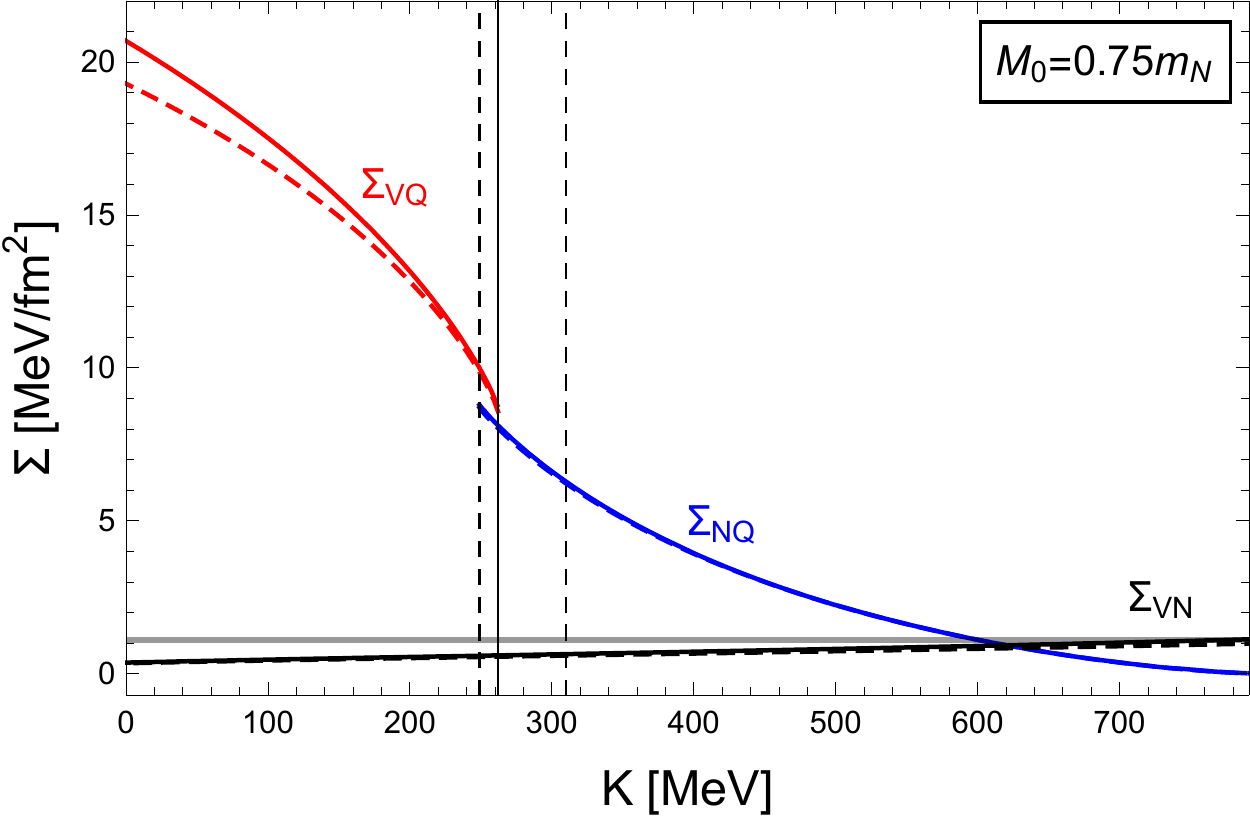}}
\hbox{\includegraphics[width=0.5\textwidth]{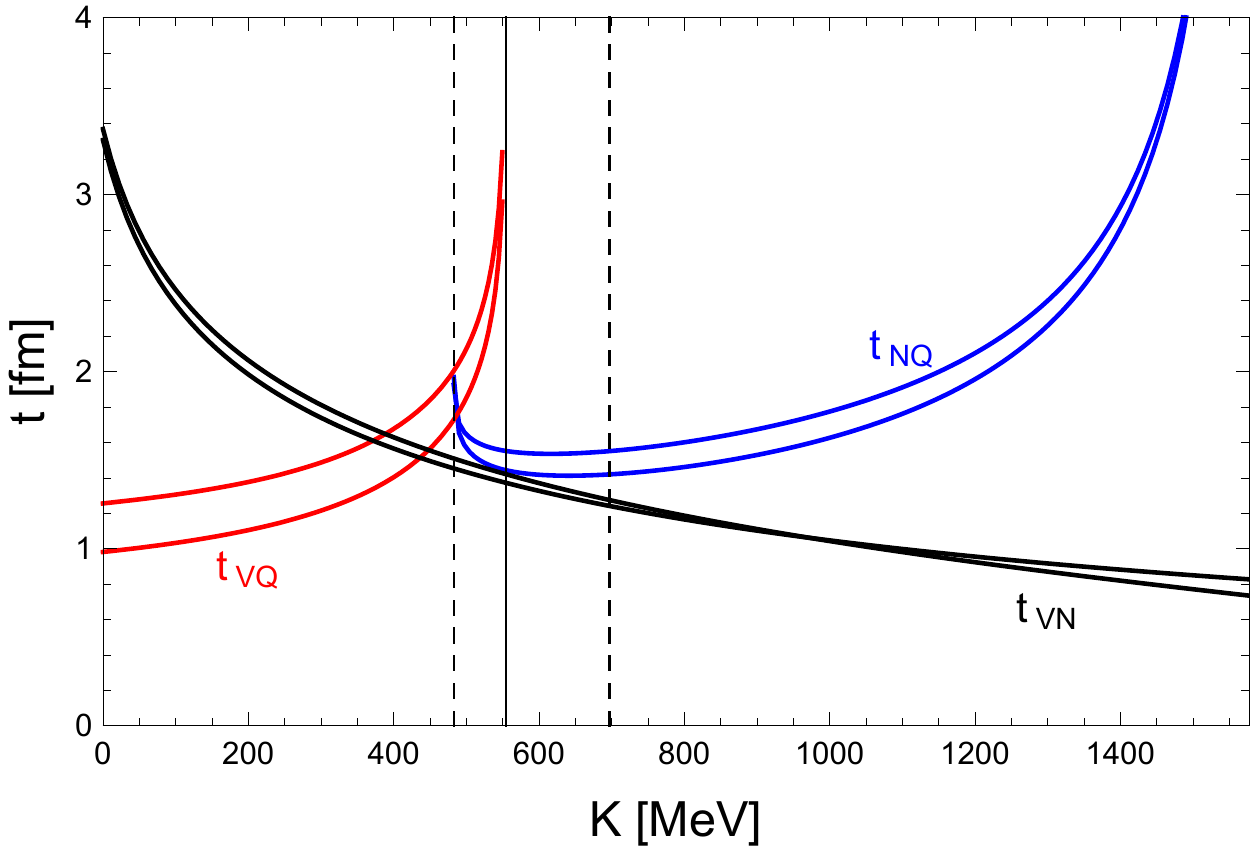}\includegraphics[width=0.5\textwidth]{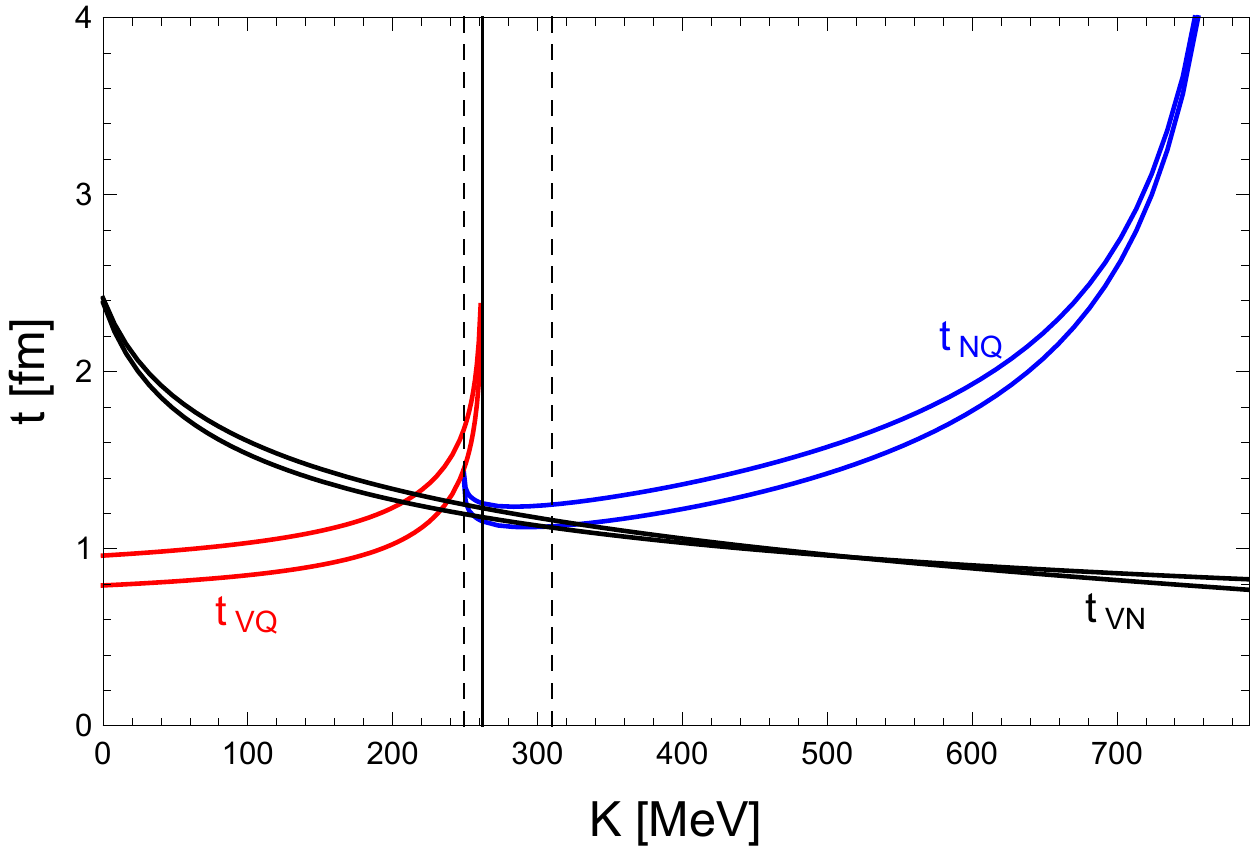}}
\caption{{\it Upper panels:} Zero-temperature surface tensions $\Sigma_{\rm VQ}$, $\Sigma_{\rm NQ}$, $\Sigma_{\rm VN}$ versus incompressibility at saturation $K$ for two different values of the effective nucleon mass at saturation $M_0$ (notice the different $K$ scales in the two panels). Solid lines are full numerical results, dashed lines are obtained from the semi-analytical approximation (\ref{SigmaApp}). The vertical dashed and solid lines correspond to the dashed and solid lines of the left panel of Fig.\ \ref{fig:paraspace}: to the left of the solid line nuclear matter only exists as a metastable state; to the left of the left dashed line N and Q never coexist, and to the right of the right dashed line V and Q never coexist. The (grey) horizontal band indicates the physical value of $\Sigma_{\rm VN}$ from Eq.\ (\ref{Sigma0}). {\it Lower panels:} Thicknesses $t_{\rm VQ}$, $t_{\rm NQ}$, $t_{\rm VN}$ of the corresponding domain walls for the same two values of $M_0$. The upper curves of each pair of curves (for $t_{\rm VN}$: upper curve for small $K$ and lower curve for large $K$) give the thickness from $\bar{\sigma}$, the lower curves give the thickness from $\bar{\omega}$. }
\label{fig:sigma2}
\end{center}
\end{figure*}

The different behavior exhibited in the right panel is best understood by consulting the upper right panel of Fig.\ \ref{fig:Mmu}. We see that, as we move from the NQ phase transition towards larger values of the chemical potential, the metastable branch of nuclear matter terminates. Therefore, this part of the spinodal region shows the usual behavior and the surface tension of the bubble goes to zero. As we move towards {\it smaller} values of the chemical potential, however, it is the stable N branch, not the metastable Q branch that terminates first. More precisely, as we go to smaller values of $\mu$, the N branch first becomes metastable itself, indicated by the solid (blue) vertical line in the right panel of Fig.\ \ref{fig:bubbles}, before this branch disappears, indicated by the left vertical dashed line. Therefore, between these two vertical lines the bubble profile interpolates between two metastable phases, while the actual stable phase is the vacuum V (i.e., one could also compute the surface tension of V bubbles immersed in Q in that region). Since the minimum of the metastable phase Q never becomes shallow in the spinodal region (this minimum only ceases to exist at a much smaller $\mu$), the surface tension remains large throughout this segment of the spinodal region.

\section{Surface tension of cold and dense matter}
\label{sec:results}

\begin{figure*} [t]
\begin{center}
\hbox{\includegraphics[width=0.5\textwidth]{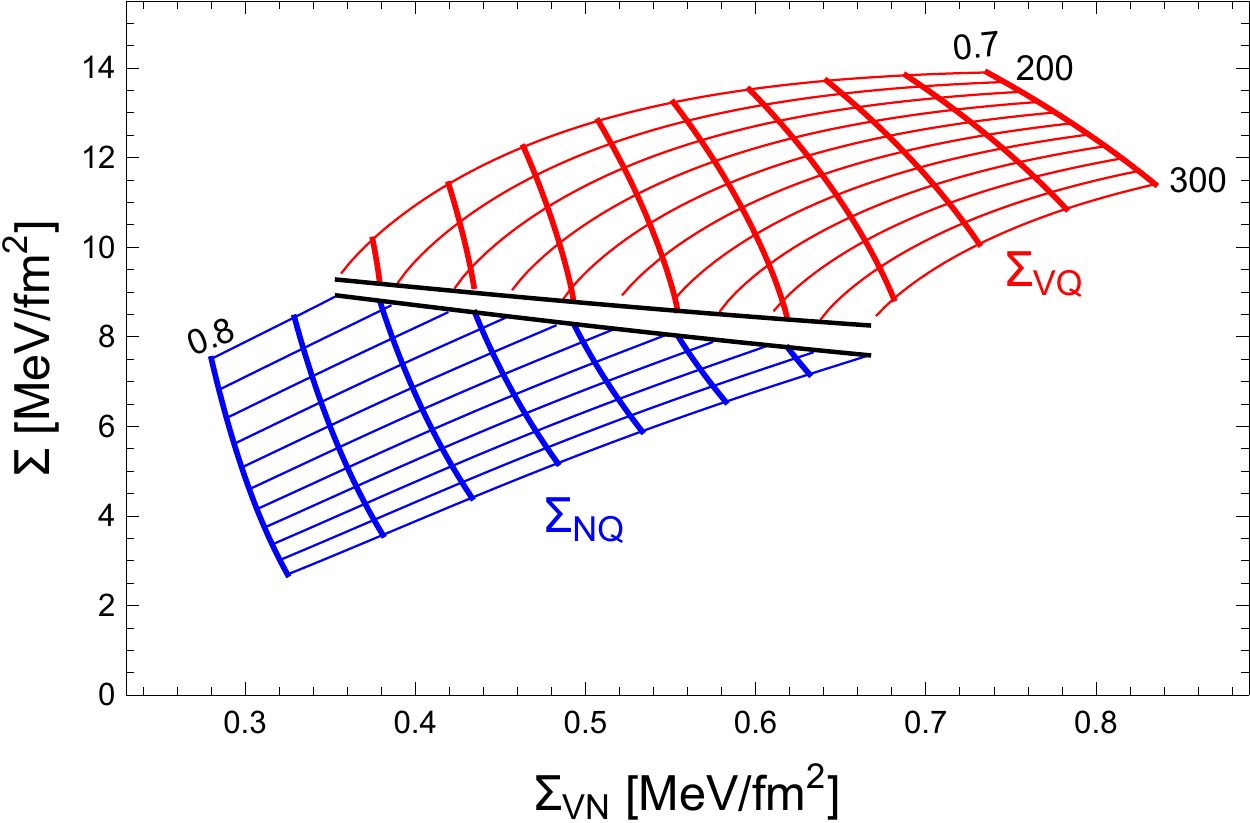}\includegraphics[width=0.5\textwidth]{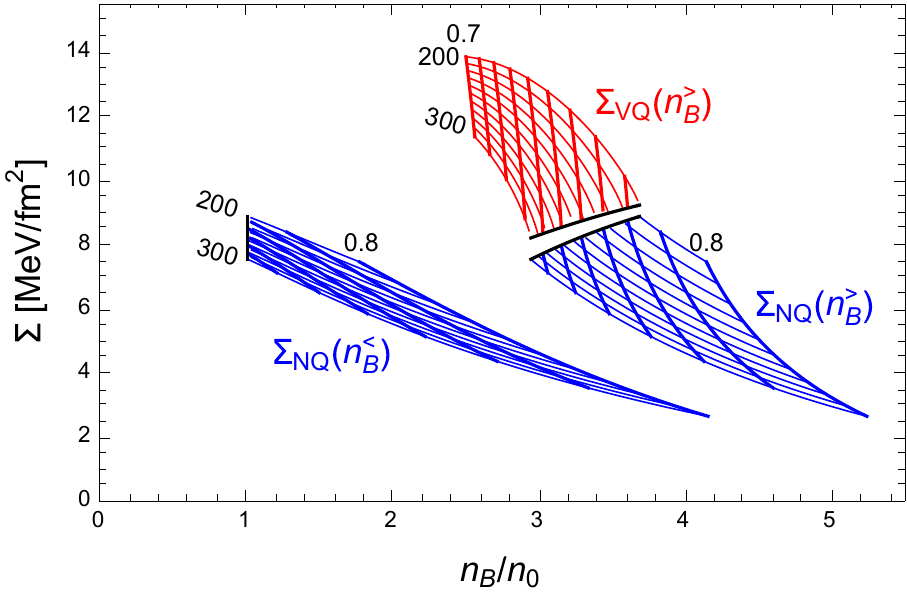}}
\caption{Zero-temperature surface tension of the nuclear-quark (NQ) and vacuum-quark (VQ) transitions versus  surface tension of the vacuum-nuclear (VN) transition (left panel) and versus baryon density just 
above, $n_B^>$, and just below, $n_B^<$, the phase transition (right panel). The results are shown for the  "allowed" parameter region for $M_0$ and $K$: thin  
lines are lines of constant $K$, from $K=200\, {\rm MeV}$ to $K=300\, {\rm MeV}$, in steps of $10\, {\rm MeV}$, thick lines are lines of constant $M_0$, from $M_0=0.7\, m_N$ to $M_0=0.8\, m_N$, in steps of 
$0.01 \, m_N$. In the parameter region where $\Sigma_{\rm VQ}$ is computed, nuclear matter only exists as a metastable state. The baryon density just below the VQ phase transition is the vacuum density $n_B^<=0$, and thus only $\Sigma_{\rm VQ}(n_B^>)$ is shown in the right panel.}
\label{fig:sigma3}
\end{center}
\end{figure*}

While in the previous section we have picked specific parameter sets to discuss temperature effects, surface tension of bubbles, and various methods to calculate the surface tension, here we show results for the surface tension in a large region of the parameter space, restricting ourselves to the surface tension of domain walls. As Fig.\ \ref{fig:bubbles} suggests, the surface tension of finite bubbles is not expected to differ significantly from the one of infinite bubbles (domain walls), except for the fact that it can become arbitrarily small at the edges of the spinodal region.

We plot the surface tension for all three possible first-order phase transitions in Fig.\ \ref{fig:sigma2} for two values of the effective nucleon mass at saturation, $M_0$, as a function of the incompressibility at saturation, $K$. The parameters of the model are then, at each point, adjusted such that saturation density $n_0$ and nuclear binding energy $E_{\rm bind}$ are held fixed  to their physical values, as explained in Sec.\ \ref{sec:fix}. In both panels the values for $K$ ranges from zero to the point where the transition between nuclear matter N and the chirally restored phase Q turns into a crossover. Therefore, $\Sigma_{\rm NQ}$ goes to zero at the upper end of the $K$ scale. The surface tensions are computed everywhere where they are defined, i.e. when the curves end there is no longer a first-order phase transition between the respective phases (see discussion of Fig.\ \ref{fig:paraspace} in Sec.\ \ref{sec:hom}). We have included results from the full numerical calculation (solid lines)
as well as from the semi-analytical approximation from Eq.\ (\ref{SigmaApp}) (dashed lines). As already suggested from the selected results in the previous section, the approximation (\ref{SigmaApp}) is in very good agreement with the full result.

Fig.\ \ref{fig:sigma2} also shows the thicknesses of the corresponding domain walls. We have chosen to plot the "90-10 thickness" $t$, defined as the range in which the condensate rises from 10\% to 90\% of the total change 
from one side of the wall to the other,
$t = x[\bar{\sigma}=\bar{\sigma}_-+0.9(\bar{\sigma}_+-\bar{\sigma}_-)]-x[\bar{\sigma}=\bar{\sigma}_-+0.1(\bar{\sigma}_+-\bar{\sigma}_-)]$. The same quantity can be computed for the 
second condensate $\bar{\omega}$, with a slightly different result. We show both results in the lower panels of Fig.\ \ref{fig:sigma2} and observe that the thickness is typically in the range $t\sim (1-3)\, {\rm fm}$ for all
transitions throughout the parameter space.

In Fig.\ \ref{fig:sigma3} we plot the surface tensions in a different way, scanning the $M_0$-$K$ parameter space more systematically, but keeping the parameters constrained to the physical regime given by Eq.\ (\ref{n0E}). For simplicity, we have used the semi-analytical approximation (\ref{SigmaApp}) for this figure, because computing all curves with the full numerical procedure would have been very laborious. Since we have seen that this approximation is very good, the full result is not expected to differ much (in most parts of the plot it would be indistinguishable by naked eye from the approximation). In the left panel, we see that the "allowed" shaded rectangle from the left panel of Fig.\ \ref{fig:paraspace} is visible in a distorted way. In addition, it has acquired a "scar". This scar appears by calculating the surface tensions of both the VQ and NQ transitions 
at the point where all three phases V, N, Q have the same free energy [lower (black) solid line in the left panel of Fig.\ \ref{fig:paraspace} and tricritical point in the right panel of Fig.\ \ref{fig:paraspace}]. There is no reason why the VQ and NQ surface tensions should be identical at this point, thus the distorted rectangle is broken. In the right panel, we have plotted the surface tension versus the baryon density 
just above and just below the phase transition. This plot thus allows us to read off the location of the phase transition in density, the size of the jump in density, and the corresponding surface tension. In the case of the NQ transition, we see that the baryon density below the phase transition can become arbitrarily close to the saturation density. This is just another way of saying that for a sizable part of the allowed rectangle in $M_0$-$K$ space there is no NQ transition, but rather a direct transition from the vacuum to quark matter. At zero temperature, this VQ transition obviously occurs at zero density and thus $\Sigma_{\rm VQ}(n_B^<)$ would simply be a line sitting on the vertical axis of the right panel, which is not shown.

From Figs.\ \ref{fig:sigma2} and \ref{fig:sigma3} we draw the following conclusions:

\begin{itemize}
\item The surface tension between vacuum and quark matter is larger than that between nuclear and quark matter. This is easy to understand already from Fig.\ \ref{fig:Mmu}: the jump in the effective nucleon mass $M$ -- in fact the jump in both condensates 
$\bar{\sigma}$, $\bar{\omega}$ -- is larger for the VQ transition than for the NQ transition. In a domain wall (or a large bubble) this jump has to be bridged and thus a large difference in condensates inevitably leads to large gradients and/or a wide wall and thus to a large surface tension. 

\item The constraint given by Eq. (\ref{Sigma0}) for $\Sigma_{\rm VN}$ cannot be fulfilled simultaneously with the constraints in Eq. (\ref{n0E}) for $K$ and $M_0$, as the left panel of Fig.\ \ref{fig:sigma3} shows. In the left panel of Fig.\ \ref{fig:sigma2} we see that smaller values of $M_0$ -- stretching the allowed region somewhat, but perhaps still consistent with experiment \cite{Jaminon:1989wj,Furnstahl:1997tk} --  are needed to reach the correct value for the vacuum-nuclear surface tension. 
In this parameter regime, nuclear matter is metastable. 

\item At fixed $M_0$, the surface tension of the chiral transitions, VQ and NQ, {\it decreases} if nuclear matter at saturation is made stiffer (increasing $K$). In contrast, the surface tension of the 
VN transition {\it increases}. The surface tension of the chiral transition tends to be smaller if the transition occurs at larger baryon densities. 

\item For the surface tensions of the chiral transitions 
within the physical values of $K$ and $M_0$ we find $\Sigma_{\rm VQ}\lesssim 15\, {\rm MeV}/{\rm fm}^2 \sim (80\, {\rm MeV})^3 $ and 
$\Sigma_{\rm NQ}\lesssim 10\, {\rm MeV}/{\rm fm}^2\sim (70\, {\rm MeV})^3$. These maximal values become larger if we allow for softer nuclear matter, but not by much: even in the extreme limit $K\to 0 $, and using $M_0=0.75 \, m_N$, the maximum value is only slightly larger,  
$\Sigma_{\rm VQ} \simeq 20\, {\rm MeV}/{\rm fm}^2\sim (90\, {\rm MeV})^3$. These values are in agreement with several  
existing calculations that use some model description of quark matter, but do not contain realistic nuclear matter (such as the Nambu-Jona-Lasinio model or quark-meson models) \cite{PhysRevD.30.2379,PhysRevC.35.213,Palhares:2010be,Palhares:2011jd,PhysRevD.84.036011,Pinto:2012aq,Mintz:2012mz,Ke:2013wga,Gao:2016hks}. They are therefore  complementary to our approach, which does contain nuclear matter but only a toy version of quark matter. There are also studies that predict larger surface tensions, up to about $100\, {\rm MeV}/{\rm fm}^2$ or even larger,  
but they either do not use a single model for the two phases at the phase transition \cite{Lugones:2011xv,Lugones:2013ema,Pais:2016dng} (and, additionally, a simple approximation for the surface tension \cite{BALIAN1970401,PhysRevD.50.3328}), or they are based on simple estimates from dimensional analysis, not on a microscopic calculation \cite{PhysRevLett.70.1355,Alford:2001zr}.

\end{itemize}

\section{Summary and outlook}
\label{sec:summary}

We have calculated the surface tension of dense matter within a nucleon-meson model, which 
accounts for realistic nuclear matter, but does not contain quark degrees of freedom. We have investigated the parameter space of the model systematically to locate the possible first-order phase transitions. In doing so, we have matched the parameters to properties of nuclear matter at saturation, which defines the allowed window in the parameter space. 

First-order transitions can occur between the vacuum, nuclear matter, and quark matter (more precisely, the approximately chirally symmetric phase).
In particular, we have identified the parameter regime where there is a direct transition between the vacuum and quark matter, i.e. where nuclear matter is metastable. In this regime, we have computed the vacuum-quark surface tension, which assumes values of $\Sigma_{\rm VQ} \simeq (8-15)  \, {\rm MeV}/{\rm fm}^2$. In the remaining parameter space we have calculated the nuclear-quark surface tension, resulting in somewhat smaller values, $\Sigma_{\rm NQ} \simeq (2-10)  \, {\rm MeV}/{\rm fm}^2$. These relatively small values seem to favor the formation of quark matter in the core of neutron stars via nucleation during their formation in a supernova explosion \cite{Mintz:2009ay,Mintz:2010mh}.

We have also discussed in detail various methods of calculating the surface tension. Besides a numerical evaluation of domain wall and bubble profiles and the resulting surface tension from the free energy, we have discussed two approximations. The first one, which we called "semi-analytical approximation", reduces the calculation of the surface tension 
to a numerical integral, such that no differential equation has to be solved. In the case of a single condensate, this reduction is exact. In the case of more than one condensate (in the present setup there are two) the reduction 
involves a small-gradient approximation. We have shown that this approximation is in very good agreement with the full result. The second approximation, which we called "one-condensate approximation", was employed for the surface tension of bubbles. In this case, we have approximated the problem by a single-condensate system, such that a numerical calculation of the bubble profiles can be done more easily. This approximation has turned out to be slightly worse: in some cases 10\% off the exact value if extrapolated to infinitely large bubbles.

There are several extensions and applications of our study. We have pointed out that the spinodal regions of the first-order phase transitions can overlap, which gives rise to unconventional profiles of domain walls and bubbles. We have not investigated these possibilities systematically, and it would be interesting to see whether they may have observable consequences. Moreover, we have evaluated the model in the 
mean-field and no-sea approximations, and have employed the Thomas-Fermi approximation to calculate the surface tension. Obvious extensions would thus be to go beyond any of these approximations. One should keep in mind, however, that our model is of phenomenological nature, and thus can only give us a rough idea of the surface tension of dense QCD matter (provided there is a first-order chiral phase transition in dense QCD), even if improvements of the present approximations are performed. It would, of course, be interesting to calculate the surface tension in a model that shows a first-order chiral (or deconfinement) transition while describing both nucleons {\it and} quark degrees of freedom. However, it is very difficult to construct such a model (see for instance recent progress along these lines in holographic studies \cite{Li:2015uea,Preis:2016fsp,Preis:2016gvf}).

Although we have restricted ourselves to a theoretical calculation of the surface tension, there are obvious applications in astrophysics, where cold and dense matter can be found in neutron stars. 
Neutron star matter is highly isospin-asymmetric due to the conditions of beta-equilibrium and electric charge neutrality. Our isospin-symmetric approach can straightforwardly be extended to include these conditions.  For applications to supernova explosions it is interesting to compute the temperature dependence of the surface tension more systematically and to estimate the associated nucleation times. It would also be interesting to apply our results to a first-order transition during a neutron star merger, possibly affecting the gravitational wave signal.

\vspace{1cm}

\begin{acknowledgments}
We thank M.\ Alford, J.\ Lattimer, and D.\ Rischke for helpful comments and discussions.  E.S.F. and M.H. are partially supported by CNPq, FAPERJ and INCT-FNA Proc.\ No.\ 464898/2014-5. 
E.S.F. is partially supported by CAPES, Finance Code 001.
M. H. is also supported by CAPES Proc.\ No.\ 88881.133995/2016-01 and  88887.185090/2018-00 (via INCT-FNA), as well as FAPESP Proc.\ No.\ 2018/07833-1 and 2017/05685-2. A.S.\ is supported by the Science \& Technology Facilities Council (STFC) in the form of an Ernest Rutherford Fellowship.
\end{acknowledgments}

\appendix

\bibliography{references}

\end{document}